%% file: oopsla18-serverless-ifc.tex
\documentclass[acmlarge]{acmart}
\renewcommand\footnotetextcopyrightpermission[1]{} % removes footnote with conference information in first column
\renewcommand\footnotetextauthorsaddresses[1]{} 
\setcopyright{none}
\usepackage{wrapfig}
\usepackage{amsmath}
\usepackage{amssymb}
\usepackage{amsfonts}
\usepackage{booktabs}   %% For formal tables:
\usepackage{subcaption} %% For complex figures with subfigures/subcaptions
\usepackage{xtab}
\usepackage{multirow}

\usepackage[final,formats]{listings}
\lstnewenvironment{pseudo}[1][]
{\lstset{
    escapeinside={(*@}{@*)},
    basicstyle=\ttfamily\small,
    keywordstyle=\bfseries,
    keywordstyle=\bfseries,
    sensitive=false,
    morekeywords={if, for, fork, try, except, function, diverge},
    identifierstyle=, 
    commentstyle=\slshape, 
    stringstyle=, showstringspaces=false,
    sensitive=true,
    morecomment=[s]{/*}{*/},
    morecomment=[l]{//},
    numberstyle=\tiny,
    stepnumber=1,
    numbers=left,
    numbersep=3pt,
    emphstyle=\bfseries,
    belowskip=0pt,
    aboveskip=0pt,
    #1
}}{}

\input{macros}

\begin{document}

%% Title information
\title[Serverless IFC]{Secure Serverless Computing Using Dynamic
Information Flow Control}%\\\small{(Under review; please do not
%circulate)}}
                                        %% [Short Title] is optional;
                                        %% when present, will be used in
                                        %% header instead of Full Title.
% \titlenote{with title note}             %% \titlenote is optional;
                                        %% can be repeated if necessary;
                                        %% contents suppressed with 'anonymous'
% \subtitle{Subtitle}                     %% \subtitle is optional
% \subtitlenote{with subtitle note}       %% \subtitlenote is optional;
                                        %% can be repeated if necessary;
                                        %% contents suppressed with 'anonymous'

%% Author information
%% Contents and number of authors suppressed with 'anonymous'.
%% Each author should be introduced by \author, followed by
%% \authornote (optional), \orcid (optional), \affiliation, and
%% \email.
%% An author may have multiple affiliations and/or emails; repeat the
%% appropriate command.
%% Many elements are not rendered, but should be provided for metadata
%% extraction tools.

%% Author with single affiliation.
\author{Kalev Alpernas}
\authornote{Work done while visiting VMware Research.}          %% \authornote is optional;
%                                        %% can be repeated if necessary
%\orcid{nnnn-nnnn-nnnn-nnnn}             %% \orcid is optional
% \affiliation{
%   \institution{VMware Research}            %% \institution is required
% }
\affiliation{
  \institution{Tel Aviv University}            %% \institution is required
}
%\email{kalevalp@post.tau.ac.il}          %% \email is recommended

\author{Cormac Flanagan}
\affiliation{
  \institution{UC Santa Cruz}
}

\author{Sadjad Fouladi}
\affiliation{
  \institution{Stanford University}
}

\author{Leonid Ryzhyk}
\affiliation{
  \institution{VMware Research}
}

\author{Mooly Sagiv}
\authornotemark[1]
\affiliation{
  \institution{Tel Aviv University}            %% \institution is required
}
% \affiliation{
%   \institution{VMware Research}            %% \institution is required
% }

\author{Thomas Schmitz}
\affiliation{
  \institution{UC Santa Cruz}            %% \institution is required
}

\author{Keith Winstein}
\affiliation{
  \institution{Stanford University}
}

%%% Author with two affiliations and emails.
%\author{First2 Last2}
%\authornote{with author2 note}          %% \authornote is optional;
%                                        %% can be repeated if necessary
%\orcid{nnnn-nnnn-nnnn-nnnn}             %% \orcid is optional
%\affiliation{
%  \position{Position2a}
%  \department{Department2a}             %% \department is recommended
%  \institution{Institution2a}           %% \institution is required
%  \streetaddress{Street2a Address2a}
%  \city{City2a}
%  \state{State2a}
%  \postcode{Post-Code2a}
%  \country{Country2a}                   %% \country is recommended
%}
%\email{first2.last2@inst2a.com}         %% \email is recommended
%\affiliation{
%  \position{Position2b}
%  \department{Department2b}             %% \department is recommended
%  \institution{Institution2b}           %% \institution is required
%  \streetaddress{Street3b Address2b}
%  \city{City2b}
%  \state{State2b}
%  \postcode{Post-Code2b}
%  \country{Country2b}                   %% \country is recommended
%}
%\email{first2.last2@inst2b.org}         %% \email is recommended

%% Abstract
\input{abstract}

%% 2012 ACM Computing Classification System (CSS) concepts
%% Generate at 'http://dl.acm.org/ccs/ccs.cfm'.
% <concept_id>10011007.10011006.10011008</concept_id>
% <concept_desc>Software and its engineering~General programming languages</concept_desc>
% <concept_id>10003456.10003457.10003521.10003525</concept_id>
% <concept_desc>Social and professional topics~History of programming languages</concept_desc>
% <concept_significance>300</concept_significance>

% \ccsdesc[500]{Software and its engineering~General programming languages}
% \ccsdesc[300]{Social and professional topics~History of programming languages}
% %% End of generated code

% %% Keywords
% %% comma separated list
% \keywords{keyword1, keyword2, keyword3}  %% \keywords are mandatory in final camera-ready submission

%% \maketitle
%% Note: \maketitle command must come after title commands, author
%% commands, abstract environment, Computing Classification System
%% environment and commands, and keywords command.
\maketitle

\input{intro}
\input{motivation}
\input{design}
\input{proof}
\input{evaluation}

\input{related}
\input{conclusion}

%% Acknowledgments
\input{acks}

%% Bibliography
%\newpage  % to appease some annoying LaTeX compiler errors
%          % please remove \newpage eventually
\bibliography{refs}

%% Appendix
\appendix

\input{appendix_proof}

\end{document}

%% file: macros.tex
\usepackage{xspace}

\newcommand{\mypara}[1]{\vspace{1mm}\noindent\textit{\textbf{#1}~~~}}

\newcommand\system{Trapeze\xspace}
\newcommand{\sysgg}{\texttt{gg}\xspace}
\newcommand{\syshello}{Hello, Retail!\xspace}
\newcommand{\sysimageext}{Image Feature Extraction\xspace}
\newcommand\src[1]{\texttt{#1}}

\newcommand{\startproof}{\begin{xtabular}{@{}l@{$\quad$}l@{$\quad$}l}}
\newcommand{\finishproof}{\end{xtabular}}

\newcommand{\storeupdate}[3]{#1[#2\,{:}{=}\,#3]}

\newlength{\longrightarrowlength}
\newlength{\notlength}
\newcommand{\startrules}{\begin{array}{l@{~{,}~}lcl@{~{,}~}l|ll}}
\newcommand{\finishrules}{\end{array}}
\newcommand{\rn}[1]{\text{\textsc{[#1]}}}

\newcommand{\steprule}[5]{%
      \multicolumn{2}{l}{ #3 }
    &
      {#4}
    &
      \multicolumn{2}{l|}{ #5 }
    &
      #2
    &
      \rn{#1}
  \\%
}

\newcommand{\ssteparrow}[1]{\overset{#1}{\longrightarrow}}
\newcommand{\sstepstararrow}[1]{\overset{#1}{\longrightarrow^*}}
\newcommand{\sstep}[3]{#2\ssteparrow{#1}#3}
\newcommand{\sstepstar}[3]{#2\sstepstararrow{#1}#3}
\newcommand{\ssteprule}[5]{\steprule{#1}{#2}{#4}{\ssteparrow{#3}}{#5}}
\newcommand{\sstepstarrule}[5]{\steprule{#1}{#2}{#4}{\sstepstararrow{#3}}{#5}}

\newcommand{\hole}{{\bullet}}

\newcommand{\replace}[3]{\storeupdate{#1}{#2}{#3}}
\newcommand{\upd}[3]{\replace{#1}{#2}{#3}}

\newcommand{\letrec}[3]{\mu x{:}T.t}

\newcommand{\s}[1]{\text{\textit{#1}}}

\newcommand{\startsyntax}{\begin{array}{rclrll}}
\newcommand{\sepsyntax}{\\&&&|&}
\newcommand{\finishsyntax}{\end{array}}

\newcommand{\process}[2]{(#1,#2)}
\newcommand{\opread}[2]{\textsf{read}~#1~#2}
\newcommand{\opwrite}[3]{\textsf{write}~#1~#2~#3}
\newcommand{\opsend}[3]{\textsf{send}~#1~#2~#3}
\newcommand{\opfork}[2]{\textsf{fork}~#1~#2}
\newcommand{\opstuck}{\textsf{stop}}
\newcommand{\opraiselabel}[2]{\textsf{raiseLabel}~#1~#2}
\newcommand{\evstart}[1]{\textsf{start}~#1}
\newcommand{\evsend}[2]{\textsf{output}~#1~#2}
\newcommand{\evnoop}{\textsf{nop}}
\newcommand{\proj}[2]{#1{\downarrow_{#2}}}
\newcommand{\lequiv}[3]{#2\approx_{#1}#3}
\newcommand{\append}{{{}+{}}\hspace{-.25em}\mathchoice{\hspace{-.5em}}{\hspace{-.5em}}{}{}{{}+{}}}

%% file: abstract.tex
%% Abstract
\begin{abstract}

The rise of serverless computing provides an opportunity to rethink
cloud security.  We present an approach for securing serverless
systems using a novel form of dynamic information flow control (IFC).

We show that in serverless applications, the termination channel found 
in most existing IFC systems can be
arbitrarily amplified via multiple concurrent requests, necessitating a
stronger termination-sensitive non-interference guarantee, which we
achieve using a combination of static labeling of serverless processes
and dynamic faceted labeling of persistent data.

We describe our implementation of this approach on top of JavaScript
for AWS Lambda and OpenWhisk serverless platforms, and present three realistic case studies showing that
it can enforce important IFC security properties with low overhead.
    
%, with space and runtime overheads of less that 20\% in all cases.

%The advent of cloud native applications and the increasing popularity of public cloud platforms raises a host of questions regarding the privacy of sensitive information.
%
%However, the stateless nature of microservices running in a FaaS environment also presents an opportunity for designing secure applications.
%
%This paper presents a framework for designing secure FaaS applications on top of Amazon Lambda. The security guarantees are backed by an Information Flow Control system that leverages the stateless nature of the FaaS microservices (Lambdas) to guarantee a secure execution with few false alarms and low runtime overheads.
%
%Our system provides run-time guarantees of Progress Sensitive Non-Interference for FaaS application.
%We describe a new technique for dynamic information flow control in
%serverless architectures.
%Our technique ensures termination-sensitive non-inference.
%A prototype of our technique has been implemented on top of JavaScript and AWS Lambda.
%We show that it incurs low runtime overhead on realistic applications.
\end{abstract}

%% file: intro.tex
\section{Introduction}\label{sec:intro}

In May 2017, the Equifax credit reporting agency suffered a security
breach, leaking social security numbers and other personal information
of 145.5 million consumers~\cite{Forbes-17}.  The breach, which
exploited a code injection vulnerability in Apache
Struts~\cite{CVE-2017-5638}, became the latest in a series of
high-profile attacks on public and private clouds compromising
sensitive personal information of hundreds of million
users~\cite{Computerworld-09,TheRegister-11,Computerworld-14,Forbes-14,Wiki-17c,Wiki-17b,ZDNet-15,Wiki-17,Wired-16,ZDNet-16,DigitalTrends-16}.

Most of these attacks can be traced down to two types of faults:
misconfigurations and software vulnerabilities.  The former include
issues like incorrect database security
attributes~\cite{TechRepublic-17,DigitalTrends-16,PCWorld-10}, the
choice of weak authentication schemes~\cite{Computerworld-14}, or the
use of unpatched software~\cite{Forbes-17}.  The latter include code
and SQL injections, file inclusions, directory traversals,
etc.~\cite{CNET-11,ZDNet-15,ZDNet-16,Computerworld-09}.

%While seemingly trivial, these issues can be hard to detect in the
%haystack of thousands of software services and configuration knobs
%comprising an enterprise cloud.
%%Detecting these seemingly trivial issues in the haystack of thousands
%%of software services and configuration knobs comprising an enterprise
%%cloud is far from trivial.  For starters, such security audit
%%requires a complete inventory of software packages used in the
%%system, which can be infeasible in a large organization.
%
%Even a correctly configured cloud is likely to contain
%%remains vulnerable.  The complexity of the cloud software stack
%%guarantees the existence of
%numerous exploitable vulnerabilities, including code and SQL
%injections, file inclusions, directory traversals,
%etc.~\cite{CNET-11,ZDNet-15,ZDNet-16,Computerworld-09}.  The Common
%Vulnerabilities and Exposures (CVE) database registers hundreds of
%such vulnerabilities every month, many of them in popular cloud
%software~\cite{NVD}.  These vulnerabilities are typically detected months
%or years after they have been introduced, if at all.

%The \emph{Trusted Computing Base}, or TCB, of a system are the
%components of the system where a vulnerability could be exploited
%to leak sensitive information.

Simply put, the enormous \emph{Trusted Computing Base} (TCB) of modern
cloud applications makes it intractable to enforce information security
in these environments.

A promising avenue to a smaller TCB lies in the use of
\emph{information flow control (IFC)}-based
security~\cite{Denning-76-CACM,Sabelfeld-Myers-03-JSAC}.  In the
IFC world, information is protected by a \emph{global security
policy} that cannot be overridden by a misconfigured application.  The
policy explicitly and concisely captures constraints on end-to-end
information flow through the system, e.g., ``credit card numbers can
only be exposed to appropriate card associations (e.g., Visa or
MasterCard)''.
%and should not leave the system through any other external channel''.

The IFC system enforces the policy even for buggy or
malicious applications, thus removing application code and
configuration from the TCB of the cloud.  In particular, an
application that has been hijacked by a code injection attack should
not be able to bypass the enforcement mechanism.  This is in contrast
to security models based on access control lists or capabilities,
where, for instance, a compromised program running with database
administrator privileges can easily leak the entire database to a
remote attacker.

%Although previous research has pointed out the potential usefulness
%of IFC in cloud computing~\cite{}, we are not aware of any practical
%implementations of this idea.  Existing IFC techniques

%\begin{enumerate}
%    \item Information security in the cloud must be governed by a
%        \emph{global security policy} that cannot be overridden by a
%        misconfigured application.  The policy must explicitly and
%        concisely capture constraints on end-to-end information flow
%        through the cloud, e.g., ``credit card numbers can only be
%        exposed to appropriate card associations (e.g., Visa or
%        MasterCard) and should not leave the system through any other
%        external channel''.
%
%    \item \emph{Policy enforcement must be decoupled from application
%        code} and should operate correctly even for buggy or malicious
%        applications.  In particular, an application that has been
%        hijacked by a code injection attack should not be able to
%        bypass the enforcement mechanism.  This is similar to how
%        operating systems enforce their security policies even for
%        misbehaving applications.
%\end{enumerate}

%The first principle above suggests the mandatory access control (MAC)
%approach to cloud security.  It further argues that MAC models based
%on information flow control~\cite{} are best suited for enforcing
%information security.

Despite significant progress on IFC, it remains difficult to apply in real software.
Dynamic IFC
systems~\cite{Austin-Cormac-09-PLAS,Efstathopoulos:2005,Stefan-etal-11-ICFP,DeGroef-etal-12-CCS} incur high runtime overhead.
Static IFC~\cite{Sabelfeld-Myers-03-JSAC,Myers-99-POPL,Myers-Liskov-00-TOSEM,zdancewic2002programming} systems shift the costs to development time, usually via the use of type systems;
however, they restrict the style of programming, which complicates their adoption.
%IFC needs to accounts for state changes in the program which occurs frequently.
%Moreover, to ensure correctness stores need to be faceted as suggested in \cite{?}, which incurs exponential
%runtime overheads.

%% What we do
%In this paper, we provide the first realistic dynamic IFC for cloud computing.
%We observe that serverless architectures allows attackers to exploit
%the fact that the system stopped producing outputs in an arbitrary way.
%Therefore, we enforce a progress sensitive non-interference.
%In contrast, most existing works focus on progress insensitive non-interference which ignores non-termination issues
%which are less relevant in standard environments.

%
%few, if any, of the previous
%IFC systems are used in production software.  The main reason is the
%high runtime cost of tracking the flow of information through the
%program~\cite{}.  Static IFC systems reduce the performance overhead
%by encoding some of the information flow constraints in the language
%type system; however the requirement to use a non-standard language
%creates a barrier to practical adoption~\cite{}.

We demonstrate that \emph{IFC for cloud computing is not only
feasible, but can be implemented with low overhead and for essentially unmodified
applications.}
%In particular, our IFC facets writes to the database but does not need to facets store mutations.
%This leverages the stateless nature of cloud applications as explained below.
%In this paper we present \system, the first practical IFC system for
%the cloud.  In stark contrast with previous IFC systems, \system is
%simple (our complete implementation consists of a few hundred lines
%of code), light-weight, and applicable to existing unmodified
%applications, while enforcing strong security guarantees.
We achieve these properties by leveraging recent developments in cloud
computing, namely, the rise of \emph{serverless 
computing}~\cite{Sbarski-17}.  Initially popularized by Amazon's AWS 
Lambda~\cite{AWSLambda}, serverless computing is rapidly gaining 
adoption by cloud
providers~\cite{OpenWhisk,IBMCloudFunctions,AzureFunctions,GoogleCloudFunctions,FnProject} 
and
tenants~\cite{Eriksen-13,Expedia,CocaCola,iRobot,Reuters,HelloRetail,StreamAlert,Jonas-17,Fouladi-17} 
due to its key benefits: elastic scalability, ease of 
deployment, and flexible pay-per-use pricing.

It achieves these benefits by decoupling application logic from 
resource management.
In the serverless model, users express their applications as 
collections of functions triggered in response to user requests or 
calls by other functions.  A function can be written in any language 
and may request a certain runtime environment, including, e.g., 
specific versions of the Python interpreter and libraries.  However, 
the function is agnostic of where this environment is instantiated: a 
physical machine, a virtual machine or a container.  The cloud 
platform manages function placement and scheduling, automatically 
spawning new function instances on demand.  This requires application 
state to be decoupled from functions and placed in a shared data store 
(e.g., a database or a key-value store), allowing all function 
instances to access the state regardless of their physical placement 
in the cloud.

We argue that serverless computing has fundamental implications for 
cloud security.  In particular, it enables practical IFC for the 
cloud.
Our key observation is that \emph{a serverless function constitutes a
natural unit of information flow tracking}.  First, a serverless
function activation handles a single request on behalf of a specific
user and only accesses secrets related to this request.  Second, each
invocation starts from a clean state and does not get contaminated
with sensitive data from previous invocations.  Under a conservative
assumption that all secrets obtained during function execution
propagate to all its outputs, \emph{we can track the global flow of
information in the system by monitoring inputs and outputs of
functions comprising it.}

%In contrast, traditional IFC tracks program data-flow at the level of
%individual instructions.
%In contrast, in the traditional server-based model, a server handles
%requests on behalf of multiple users, with secrets belonging to all
%these users potentially scattered around its memory.  In this
%setting, IFC requires tracking program data-flow at the level of
%individual instructions, which can be prohibitively
%expensive~\cite{}.

\begin{wrapfigure}{r}{0.4\linewidth}
    \center
    \includegraphics[width=0.8\linewidth]{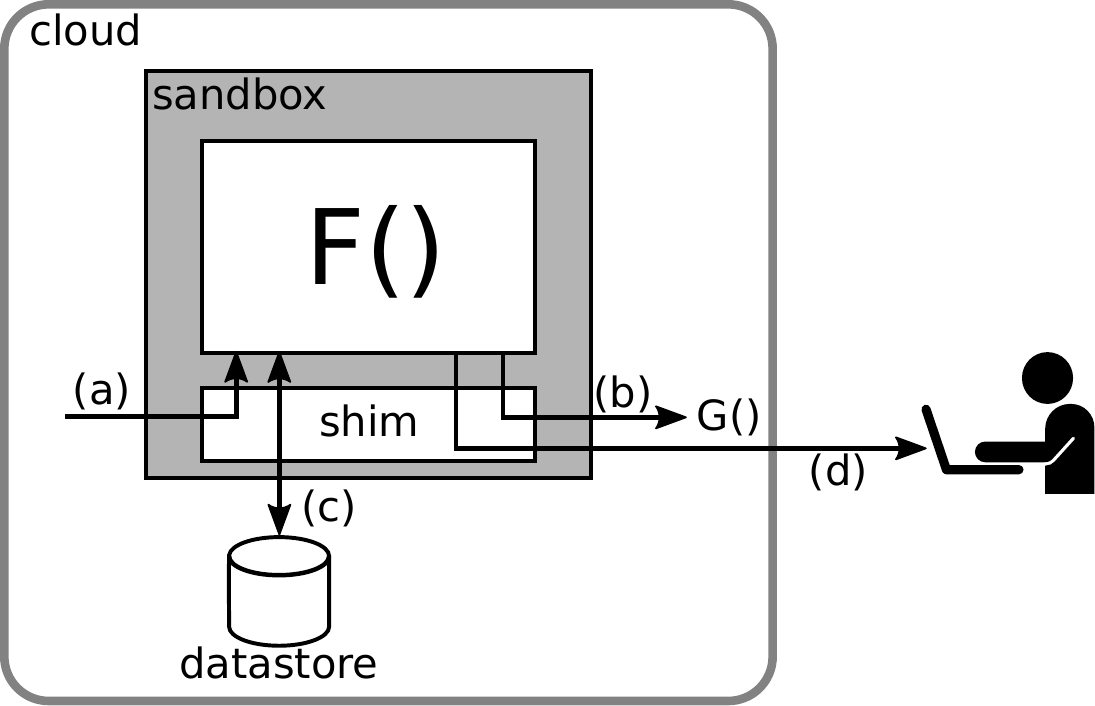}
    \caption{\system architecture. Serverless function $F$ is
    encapsulated in a sandbox.  All inputs and outputs of $F$,
    including (a) invocations of $F$ from within and from outside of
    the cloud, (b) calls to other serverless functions, (c) reads and
    writes to shared data stores, and (d) external communication
    channels, are monitored by the security shim.}\label{fig:arch}
\end{wrapfigure}

Based on this observation, we develop the first IFC system for
serverless applications, called \system.  \system encapsulates each
unmodified serverless function in a \emph{sandbox}, which intercepts
all interactions between the function and the rest of the world,
including other functions, shared data stores, and external
communication channels, and redirects them to the \emph{security shim}
(Figure~\ref{fig:arch}).  The shim tracks information flow and
enforces the global security policy.

The class of supported policies, along with policy enforcement rules, 
is defined by \system's \emph{dynamic IFC model}.  The model addresses 
a weakness in most existing static and dynamic IFC systems, which leak 
information through the \emph{termination channel}, whereby an 
adversary can infer secrets by observing termination or 
non-termination of the program.  The massively parallel nature of the 
serverless environment amplifies this weakness, allowing the attacker 
to construct a high-bandwidth information channel, effectively 
defeating the purpose of IFC (Section~\ref{sec:motivation}).

Our IFC model eliminates this channel by enforcing a strong security
property known as \emph{termination-sensitive non-interference}
(TSNI)~\cite{Sabelfeld-Sands-01}.  Only a handful of previous IFC
models achieve this level of security; however these models are either
too restrictive for most practical
applications~\cite{Sabelfeld-Sands-01,Bell-LaPadula-73,Smith-Volpano-98},
or introduce significant overhead~\cite{Devriese-Piessens-10-SSP} (see
Section~\ref{sec:related}).

\system achieves TSNI through a novel combination of static program
labeling with dynamic labeling of the data store based on a faceted
store semantics. Static program labeling restricts the sensitivity of
data a serverless function can observe ahead of time and is key to
eliminating the termination channel.  Dynamic data labeling is crucial
to securing unmodified applications that do not statically partition
the data store into security compartments, while the faceted store
semantics eliminates implicit storage channels.  We present a formal 
proof, validated using the Coq proof
assistant, that our model enforces TSNI.

We evaluate \system on three real-world serverless applications: an
online retail store~\cite{HelloRetail}, a parallel build
system~\cite{gg17}, and an image feature extraction
service~\cite{Serverless-examples}.  We use \system to secure these
applications with minimal changes to application code and with low 
runtime
overhead.

Thus, our key contributions in this work are (1) a light-weight
\emph{IFC shim architecture} for serverless computing, (2) a new
\emph{IFC model} that enforces TSNI, along with its formal semantics
and proof of correctness, and (3) an experimental evaluation of the
architecture and the model on three serverless applications.

Finally, we point out that our IFC model is not limited to the 
serverless domain.  Generally speaking, it applies to any reactive 
system that decouples computation from state.  Algorithmically, we 
show that TSNI can be enforced in such systems efficiently.  Examples 
of such systems include, e.g., Hadoop~\cite{Hadoop}, Apache Spark~\cite{Zaharia-2012} and Stateless 
Network Functions~\cite{Kablan-17}.

%It is possible to view our approach as IFC scheme for a limited class of reactive programs where only
%the mutable states is stored in a database.
%Cloud computing frameworks such as Amazon's AWS Lambda and OpenWisc nicely fits this model but our techniques
%will apply in other situations.
%Algorithmically, we show that IFC can be implemented in an efficient manner for such programs with limited overheads.
%The main idea is that only database writes have to be facted.

%% file: motivation.tex
\section{Why a new IFC model for serverless?}\label{sec:motivation}

As discussed above, a serverless function offers a convenient unit of 
information tracking, enabling practical IFC for serverless 
applications.  It seems natural that our next step should be adapting 
one of many existing IFC models to the serverless environment.  However, we do 
not take this path, as existing models do not provide adequate 
security for serverless applications.  Specifically, most previous IFC 
models, both dynamic and static, enforce a security property known as 
\emph{termination-insensitive non-interference}
(TINI)~\cite{Sabelfeld-Sands-01}.  Intuitively, TINI guarantees that an 
attacker cannot deduce secrets stored in the 
system from its non-secret outputs.  However, they may be able to 
deduce part of a secret from the fact that the system stopped 
producing outputs.  

This information channel, known as the \emph{termination channel}, 
is often disregarded because it has low bandwidth, typically leaking just 
a single bit.  This is not true in serverless systems.  Below, we 
construct an attack on a serverless application, that amplifies the 
termination channel by spawning many parallel computations, each leaking one bit.  

\begin{example}\label{ex:termination}
    Consider a serverless system with two users: a benign user Bob and 
a malicious user Eve.  We introduce security labels $b$ and $e$ to tag 
Bob's and Eve's data respectively.  
    Labels form a \emph{lattice}, with labels higher up in the lattice 
    representing more secret data.
    Bob and Eve are mutually 
distrusting, therefore their labels are incomparable (Figure~\ref{fig:lattice}).

\begin{figure}
\centering
\begin{subfigure}[b]{0.35\textwidth}
    \centering
    \includegraphics[width=0.6\linewidth]{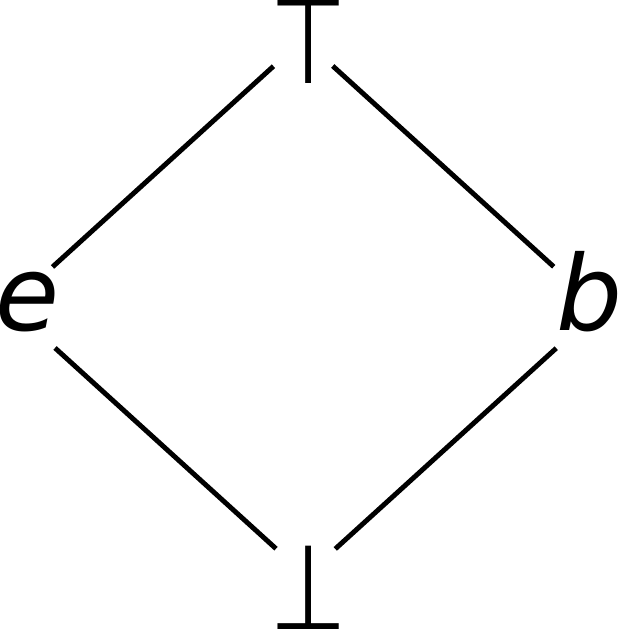}
    \caption{Example security lattice. $\top$ is the most secret label at the top of the
    lattice; $\bot$ is the bottom (least secret) 
    label.}\label{fig:lattice}
\end{subfigure}
\hspace{10mm}
\begin{subfigure}[b]{0.5\textwidth}
    \begin{pseudo}[firstnumber=1]
function F/*compromised*/():
(*@$e$@*)  for i = 0 to 63:
(*@$e$@*)    fork H(i)
(*@$e$@*)  secret = store.read(100)
(*@$\top$@*)  for i = 0 to 63:
(*@$\top$@*)    if secret[i]:
(*@$\top$@*)      store.write(i,1)

function H(i):
(*@$e$@*)    delay(1s)
(*@$e$@*)    store.read(i)
(*@$e/\top$@*)  eve.send(i) /*diverges if label is (*@$\top$@*)*/
    \end{pseudo}
    \caption{High-bandwidth termination channel in the classical 
    dynamic IFC model.}\label{fig:exploit3}
\end{subfigure}

\caption{A security lattice and a termination channel.}
\end{figure}

Eve launches a code injection attack against serverless function \src{F}, 
forcing the function to execute malicious code in Figure~\ref{fig:exploit3}.  
This code is designed to leak Bob's 
64-bit secret stored under key $100$ in the key-value store to Eve.

    We assume that the system is secured using a classical dynamic IFC 
model, where, upon reading a value from the store, the function's 
label gets bumped to the least upper bound of the function's previous 
label and the value's label (note that, since the security shim cannot 
observe the data flow inside the function, it must conservatively 
taint the entire function with this label).  The function initially 
runs with Eve's label $e$ (every line in the listing is annotated 
with the current label of the function).  Before reading Bob's secret, 
the function forks 64 instances of helper function \src{H} (lines~2--3).  Each instance of H 
will leak one bit of the secret.  Next, we read the secret, 
bumping the function's dynamic label to $b \sqcup e = \top$ and encode 
it using the first 64 keys in the store: we store an arbitrary value 
under key $i$ iff the $i$th bit of the secret is $1$ (lines~5--7).

    The helper function \src{H} also starts with initial label $e$.  
After a short delay (line~10), it reads the key equal to the 
function's argument $i$ from the store (line~11).  At this point 
the function's label either bumps to $\top$ if the corresponding bit 
of the secret is $1$ or stays at $e$ otherwise.  Line~12 attempts to 
send a message to Eve (here, \src{eve} is a handle to Eve's HTTP 
session), which succeeds if the label of the channel ($e$) is higher than or
equal to the function's current label and diverges otherwise.  
Eventually, the attacker at the other end of the channel receives the list of bits of 
the secret equal to $0$.
    \qed
\end{example}

We implemented the above attack in AWS Lambda  
and were able to leak 170~bits per second by scaling the number of 
threads.  Thus, by elastically scaling the computation, the serverless 
architecture also scales the termination channel.
%present in conventional IFC systems, essentially defeating the purpose of IFC.
This is inline with theoretical results, which suggest that in a 
concurrent system, the termination channel can leak a secret in time 
linear in the size of the secret~\cite{Askarov-08}.

We therefore aim for a stronger security guarantee, known as 
\emph{termination-sensitive non-interference} (TSNI)~\cite{Sabelfeld-Sands-01}, which eliminates 
the termination channel.  Note that the termination channel in the 
above example arises as the function's label, and hence its ability to 
send to an external channel, depends on the labels of values the 
function reads from the store.  In contrast, our proposed 
model assigns a static security label to each function 
activation.  To do so, we take advantage of the fact that a serverless 
function always runs on behalf of a specific user and can be assigned 
a corresponding security label.  The complete model, presented in 
Sections~\ref{sec:design} and~\ref{sec:formal}, also offers a secure 
way to dynamically increase the function's label without introducing a 
side channel.

The function's label determines its view of the data store: the 
function can only observe the existence of data whose label does not 
exceed the function's label.  For example, when reading a key that 
contains a secret above function's current label, the store 
returns the same result as if the key was not present in the 
store.  This information hiding semantics is somewhat tricky to 
maintain when multiple functions with incomparable labels write to the 
same store location.  We avoid information leaks 
in this situation by employing \emph{faceted store} 
semantics, where each record can contain several values (facets) with 
different security labels~\cite{Austin-Flanagan-12-POPL,Yang-16}.

To the best of our knowledge, \system is the first IFC system to 
combine static program labeling with dynamic labeling of data (using 
faceting).  This combination eliminates termination and storage 
channels and enforces a strong security property, TSNI.

%% file: design.tex
\section{Informal design}\label{sec:design}

\subsection{Threat model and assumptions}

We assume that the following entities are trusted, i.e., not malicious 
or compromised: (1) the cloud operator, (2) physical hosts and network, 
(3) system software (OS, hypervisor, container 
manager, scheduler), (4) serverless runtime, (5) shared data stores, 
(6) the sandboxing technology.  Assumptions (1) through (5) can in the 
future be relaxed with the help of a secure enclave technology such as 
Intel SGX~\cite{SGX,Hunt-16}, data encryption, and software 
verification~\cite{Denning-77,Heintze-98}.  

We further trust the serverless application administrator to enforce 
the following invariants on application configuration: (1) all data 
stores used by the application are configured to only be accessible 
from serverless functions, (2) all serverless functions in the system 
are sandboxed.  

Finally, we trust the application developer to correctly define the
application's information flow policy and declassifier functions 
(Section~\ref{sec:declassifiers}). 

The rest of the application is untrusted.  In particular, we assume 
that the attacker can compromise application code running inside 
serverless functions, including any of the frameworks and libraries it 
uses.

In this paper we focus on data \emph{confidentiality}, i.e., 
protecting sensitive data from being exposed to unauthorized users.  
The complementary problem of enforcing data \emph{integrity}, i.e., protecting 
data from unauthorized modification is outside 
the scope of \system, although it can also be enforced with the help 
of IFC techniques~\cite{Sabelfeld-Myers-03-JSAC}.  Finally, we are not 
concerned with timing and other covert channels~\cite{Biswas-17}.  An 
example of such a channel is a function running with a high label 
communicating with a function running with a low label by modulating 
its CPU or memory usage. 

\subsection{Security lattice}\label{sec:lattice}

We start the construction of our IFC model with the lattice of 
security labels.  Labels represent security classes of information 
flowing through the system.  \system does not assign any specific 
semantics to labels; however in practice they typically represent 
users or roles of the system.  

%\begin{example}[Security lattice]
%    Figure~\ref{fig:lattice} shows the security lattice that matches
%    Example~\ref{ex:termination}.
%    %our running example: we assign unique security labels $b$ and $e$ 
%    %to users Bob and Eve respectively.  Users are mutually 
%    %distrusting, therefore their labels are incomparable.
%    \qed
%\end{example}

\system relies on a trusted authentication gateway to tag all external 
input and output channels with correct security labels.  For example, 
when Eve establishes an HTTP session with the system, the session gets 
tagged with Eve's label $e$.

%Note that, while we trust the authentication mechanism, \system helps 
%protect user credentials from unauthorized access, as discussed in 
%Section~\ref{}.

Given the labeling of inputs and outputs, \system applies information 
flow rules presented below to enforce that information 
received through an input channel labeled $l_i$ can only be exposed 
through an output channel labeled $l_o$ if $l_i \sqsubseteq l_o$.

\subsection{Information flow rules}

%As introduced in Section~\ref{}, \system relies on the security shim 
%to enforce information flow rules.  

The choice of information flow rules determines two critical 
properties of an IFC system: \emph{security} and \emph{transparency}.  
The former characterizes the class of insecure behaviors the system 
prevents.  The latter characterizes the class of secure 
programs that the system executes with unmodified semantics and that
therefore do not need to be modified to work with \system.  
\system's enforces the strong security 
property of TSNI at the cost of some loss of 
transparency that, we argue, is acceptable in serverless systems.

\system assigns a runtime security label to every serverless function 
activation.  This label is derived from the event that triggered the 
function.  In particular, if the function was invoked via an HTTP 
request from a user, it obtains the user's security label.  
Alternatively, when invoked by another function, it inherits the 
caller's label. The function's label controls its ability to send to 
an output channel: a send is only allowed if the function's label 
is smaller than or equal to the channel label.

\system also dynamically labels records in the data store.  To this 
end, the security shim intercepts data store operations issued by the 
function and modifies them to insert and check security 
labels.  When a function creates or updates a record in the  
store, the record inherits the function's label (see detailed write 
semantics below).  When reading from the store, the function only 
observes values whose labels are below or equal to its own label.  
From the function's perspective, the store behaves \emph{as if} it did not
contain any data that the function may not observe.

A function can upgrade its label to an arbitrary higher label using the 
\textsf{raiseLabel} operation.  This 
operation does not introduce an unauthorized information channel, as 
the decision to upgrade cannot depend on secrets above function's 
previous label (such secrets are simply invisible to the 
function).  The upgrade mechanism is useful, for example, when a 
function running on behalf of a regular user needs to update global 
statistics on behalf of a superuser.  Upgrade is a one-way operation: 
a function's label can never be downgraded below its current value.

\begin{wrapfigure}{l}{0.4\textwidth}
    \centering
    \hspace{5mm}
    \begin{pseudo}[firstnumber=1]
function Fb():
(*@$b$@*)  secret = store.read(100)
(*@$b$@*)  for i = 0 to 63:
(*@$b$@*)    if secret[i] == 1:
(*@$b$@*)      store.write(i,1)

function Fe():
(*@$e$@*)  x = 0
(*@$e$@*)  for i = 0 to 63:
(*@$e$@*)    store.write(i,1234)
(*@$e$@*)    if store.read(i) == 1:
(*@$e$@*)       x[i] = 1
(*@$e$@*)  eve.send(x)
    \end{pseudo}
    \caption{Implicit storage channel via conflicting 
    writes.}\label{fig:exploit2}
\end{wrapfigure}

\mypara{Store semantics}
\system's security shim conceals the existence of 
data whose security label is not less than or equal to the function's 
label.  Maintaining this semantics is 
straightforward when all writes to a data store location carry the 
same label.  In the presence of conflicting writes, a resolution 
mechanism is required to avoid implicit storage 
channels~\cite{Austin-Cormac-10-PLAS}.  When the 
conflicting labels are comparable, conflict resolution can be 
performed, e.g., using the
no-sensitive-upgrade rule~\cite{Austin-Cormac-10-PLAS}.  However, no 
such mechanism that avoids storage channels exists for writes with 
incomparable labels.  The following example illustrates the problem:

\begin{example}[Implicit storage channel]
    \label{ex:implicit}
    Suppose we resolve conflicts by ignoring writes when an
    incomparable label exists in the store (alternative strategies, 
    e.g., silently overriding the existing value or failing the 
    conflicting write are similarly vulnerable).  
    Figure~\ref{fig:exploit2} shows two functions running with labels 
    $b$ and $e$ respectively that collude to leak Bob's 64-bit secret 
    to Eve.  $Fb$ reads the secret in line~2; however it does not have 
    the authority to send it to Eve directly.  Instead it encodes each 
    bit of the secret using a record in the key-value store, as in 
    Example~\ref{ex:termination} (lines~3--5).  $Fe$ reconstructs the 
    secret by attempting to write to locations $0$ to $63$ and then 
    reading the value back in.  Writes to locations that correspond to 
    $1$s are ignored, indicating that the corresponding bit of the 
    secret is $1$.
    \qed
\end{example}

We eliminate such unauthorized flows using \emph{faceted store} 
semantics, where each record can contain several values (facets) with 
different labels~\cite{Austin-Flanagan-12-POPL,Yang-16}.  
Facets are created dynamically: when a value with a new label is 
stored in the record, a facet is created for it (see 
Section~\ref{sec:formal} for precise semantics).  A read 
returns the most recent write that is visible to the function.  Thus, facets 
conceal writes with label $b$ from  
a function running with label $e$, unless $b \sqsubseteq e$.  

\begin{example}
    We replay the example in Figure~\ref{fig:exploit2} with faceted 
    store semantics.  Since function labels $b$ and $e$ are 
    incomparable, their respective writes will go into different 
    facets.  The functions do not observe each other's writes either 
    explicitly or indirectly, as in Example~\ref{ex:implicit}.
    \qed
\end{example}

Faceted stores were previously introduced in IFC research in work on 
faceted execution~\cite{Austin-Flanagan-12-POPL,Yang-16,Austin-13}.  The fundamental difference from our 
approach is in the read semantics.  Under faceted execution, a 
read conceptually forks the program, creating a 
separate branch for each facet read from the store.  If the program 
sends to an external channel, only the branch whose label is 
compatible with that of the channel is allowed to send.  
Similar to our design, faceted execution eliminates storage 
channels; however it does so at a potentially high runtime cost and 
may become impractical in a system with a large security lattice.  In 
contrast, \system avoids faceted-execution using the apriori knowledge 
of the function's label by pruning all incompatible facets at read 
time.  Existing faceted execution systems expose the termination 
channel and therefore enforce TINI, whereas \system enforces the 
stronger TSNI property.

In the practical use of \system, faceting is an exceptional situation.  
\system is designed to run unmodified applications that assume 
conventional store semantics.  The moment multiple facets are 
created in some store location, this semantics is violated, as different functions can now 
observe different values at the same location.  \system treats 
such situations as attempted exploits and notifies the administrator, 
who can then take recovery actions, e.g., remove the offending 
function from the system and rollback the store 
to the previous consistent state.  In the meanwhile \system guarantees 
that the system continues running without exposing any sensitive 
information to the attacker.

Faceted store semantics is emulated by the security shim on top of a 
conventional non-faceted store.  In Section~\ref{sec:evaluation}, we 
implement facets on top of a key-value store.  Yang et al.~\cite{Yang-16} 
present the design of a faceted SQL database.

\mypara{Transparency}
The flip side of \system's strong TSNI security guarantees and 
light-weight protection is the theoretical loss of transparency, i.e., 
the ability to run existing unmodified applications.  By assigning a 
static security label to a function, we restrict data that is visible 
to it.  In particular, the function cannot access values above its 
security level even if it does not send these values (or anything 
derived from them) through unauthorized channels.  On the other hand, 
all writes to the data store performed by the function are 
conservatively labeled with the function's label even if they do not 
carry any secrets.  

Both problems can be addressed by refactoring the application.  In 
particular, the function can gain access to secret data via the 
\textsf{raiseLabel} operation.  Conversely, one can avoid tainting 
data with excessively high labels by splitting the offending function 
into several functions that run with lower labels.  However, if many 
such changes are required in order to adapt existing applications to 
work with \system, this will create a barrier to \system's practical 
adoption.

Our evaluation in Section~\ref{sec:evaluation} indicates that in practice 
the loss of transparency is not an issue in serverless applications due
to the common serverless design practice where every function only accesses values that are 
related to a specific small task and are therefore likely to have compatible 
security labels.

\subsection{Declassifiers}\label{sec:declassifiers}

Many real-world applications allow limited flow of information down 
the security lattice.  For example, a credit reporting agency may 
make the distribution of consumers across credit score bands publicly 
available.  This statistics is computed based on the credit history of 
all consumers and must therefore be labeled with the least upper bound 
of all their labels.  However, since the aggregate statistics exposes 
negligible amount of information about individual consumers, it can be 
safely \emph{declassified}.

Similar to previous IFC models, \system introduces 
\emph{declassifiers} to support such scenarios.  A declassifier is a 
triple $\langle h, l, D \rangle$, where $h$ and $l$ are security 
labels, such that $l \sqsubset h$, and $D$ is a serverless function.  
A declassifier is invoked just like any other serverless function; 
however its security label is computed using special rules.  Let $x$ 
be the label of the calling function.  Then the declassifier is 
assigned label $x'$, such that:
$$
x' = \begin{cases}
        l, & \mbox{if } l \sqsubseteq x \sqsubseteq h\\
        x, & \mbox{otherwise}
     \end{cases}
$$

By design, declassifiers violate the non-interference property; 
therefore the formal model and proofs in Section~\ref{sec:formal} are 
given for the pure IFC model without declassifiers.

%% file: proof.tex
\section{Formal semantics}\label{sec:formal}

\begin{figure*}
\begin{footnotesize}
$\begin{array}{cc}
\begin{minipage}[t]{0.5\textwidth}
$\startsyntax
  l     & \in & \s{Label}  \\
  v     & \in & \s{Value}  \\
  S     & \in & \s{LabeledValueSeq} & = & (\s{Value}\times\s{Label})^*  \\
  k     & \in & \s{Key}  \\
  \sigma& \in & \s{Store} & = & \s{Key}\to\s{LabeledValueSeq}  \\
  t     & \in & \s{Thread}  \\
  p     & \in & \s{Process} & ::= & \process{t}{l}  \\
  ps    & \in & \s{Processes} & = & \text{multisets of processes}  \\
  \Sigma& \in & \s{State} & ::= & (\sigma,ps)  \\
\finishsyntax$
\end{minipage}
&
\begin{minipage}[t]{0.5\textwidth}
$\startsyntax
  e     & \in & \s{Event} & ::= &
      \evstart{p}
    \sepsyntax
      \evsend{ch}{v}
    \sepsyntax
      \evnoop
    \\
  ch    & \in & \s{OutputChannel}  \\
  es    & \in & \s{Events} & = & \s{Event}^*  \\
  op    & \in & \s{Operation} & ::= &
      \opread{k}{f}
    \sepsyntax
      \opwrite{k}{v}{t}
    \sepsyntax
      \opsend{ch}{v}{t}
    \sepsyntax
      \opfork{t}{t}
    \sepsyntax
      \opraiselabel{l}{t}
    \sepsyntax
      \opstuck
    \\
  f     & \in & \s{ReadContinuation} & = & (\s{Value}\times\s{Label})_\bot\to\s{Thread}  \\
\finishsyntax$
\end{minipage}
\\
\\
\text{(a) Serverless execution state.}
&
\text{(b) External events and I/O operations.}
\\
~
\end{array}$
\\
$\startrules
\multicolumn{6}{@{}l}{
  \text{\fbox{$\sstep{e}{\Sigma}{\Sigma}$} State transition relation}
}\\
  \ssteprule{s-start}
        {}
        {\evstart{p}}
        {(\sigma,ps)}
        {(\sigma,ps\uplus\{p\})}
  \ssteprule{s-send}
        {\s{run}(t)=\opsend{ch}{v}{t'}\qquad l\sqsubseteq\s{label}(ch)}
        {\evsend{ch}{v}}
        {(\sigma,ps\uplus\{\process{t}{l}\})}
        {(\sigma,ps\uplus\{\process{t'}{l}\})}
  \ssteprule{s-read}
        {\s{run}(t)=\opread{k}{f}}
        {\evnoop}
        {(\sigma,ps\uplus\{\process{t}{l}\})}
        {(\sigma,ps\uplus\{\process{f(\s{last}(\proj{\sigma(k)}{l}))}{l}\})}
  \ssteprule{s-write}
        {\s{run}(t)=\opwrite{k}{v}{t'}}
        {\evnoop}
        {(\sigma,ps\uplus\{\process{t}{l}\})}
        {(\upd{\sigma}{k}{\s{write}(\sigma(k),v,l)},ps\uplus\{\process{t'}{l}\})}
  \ssteprule{s-fork}
        {\s{run}(t) = \opfork{t'}{t''}}
        {\evnoop}
        {(\sigma,ps\uplus\{\process{t}{l}\})}
        {(\sigma,ps\uplus\{\process{t'}{l}\}\uplus\{\process{t''}{l}\})}
  \ssteprule{s-raise-label}
        {\s{run}(t) = \opraiselabel{l'}{t'}\qquad l\sqsubseteq l'}
        {\evnoop}
        {(\sigma,ps\uplus\{\process{t}{l}\})}
        {(\sigma,ps\uplus\{\process{t'}{l'}\})}
  \ssteprule{s-skip}
        {}
        {\evnoop}
        {(\sigma,ps)}
        {(\sigma,ps)}
\multicolumn{6}{@{}l}{
  \text{\fbox{$\sstepstar{es}{\Sigma}{\Sigma}$} Multiple-step state transition relation}
}\\
  \sstepstarrule{refl}
        {}
        {\epsilon}
        {\Sigma}
        {\Sigma}
  \sstepstarrule{trans}
        {
            \sstep{e}{\Sigma}{\Sigma'}
          \qquad
            \sstepstar{es}{\Sigma'}{\Sigma''}
        }
        {e\append es}
        {\Sigma}
        {\Sigma''}
\finishrules$
\\~\\
(c) Transition relations.
\caption{Formal semantics.}
\label{formal_semantics}
\end{footnotesize}
\end{figure*}

In this section, we formalize our IFC semantics for serverless systems with an underlying persistent key-value store.
Since computations at different security labels might write to the same key, the key-value store maps each key to a \textit{set} of values (facets), each with a different label, and we order this set into a sequence according to the temporal order of the writes.
The initial store $\sigma_0$ maps each key to the empty sequence: $\sigma_0=\lambda k.\epsilon$.

A state $\Sigma\in\s{State}$ of the system consists of a key-value store $\sigma\in\s{Store}$, plus a multiset of currently executing serverless function activations called \textit{processes} (see Figure \ref{formal_semantics}(a)).
Each process $p=\process{t}{l}$ consists of a \textit{thread} $t$ plus its associated security label $l\in\s{Label}$.

Observable events $e\in\s{Event}$ of the system include the input event $\evstart{p}$, which starts a new process $p$,\footnote{
  Serverless systems spawn a new function/process $p$ to handle each incoming event. Here, we assume that each incoming event contains that new process, to simplify the formal development.
} and the output event $\evsend{ch}{v}$, which sends the value $v$ on output channel $ch$ (see Figure \ref{formal_semantics}(b)).

\begin{figure}%{r}{0.5\linewidth}
$\begin{array}{@{}r@{~}c@{~}l@{}}
%  &           & \proj{\hole}{\hole} & : & \text{``Various things''}\to\s{Label}\to\text{``Various things''}  \\
  \proj{S}{l} & = & S\setminus\{(v',l')\in S~|~l'\not\sqsubseteq l\}  \\
  \proj{ps}{l} & = & \text{multiset}\{\process{t}{l'}\in ps~|~l'\sqsubseteq l\}  \\
  \proj{\sigma}{l} & = & \lambda k. \proj{\sigma(k)}{l}  \\
  \proj{e}{l} & = &
    \begin{cases}
      e  & \exists t,l'.~e=\evstart{\process{t}{l'}}\text{ and }l'\sqsubseteq l  \\
      e  & \exists ch,v.~e=\evsend{ch}{v}\text{ and }\s{label}(ch)\sqsubseteq l  \\
      \evnoop & \text{otherwise.}
    \end{cases}  \\
  \proj{(e_1\ldots e_n)}{l} & = & \proj{e_1}{l}\ldots\proj{e_n}{l}  \\
  \proj{(\sigma,ps)}{l} & = & (\proj{\sigma}{l},\proj{ps}{l})   \\
\finishsyntax$
\caption{Definition of projection function.}
\label{definition_of_projection}
\end{figure}

The state transition relation $\sstep{e}{\Sigma}{\Sigma'}$ shown in Figure \ref{formal_semantics}(c) describes how the system executes.
The first rule $\rn{s-start}$ handles an incoming event $\evstart{p}$ simply by adding $p$ to the multiset of processes.
The next five transition rules all involve executing a particular process $\process{t}{l}$ until its next I/O operation $op\in\s{Operation}$.
For maximal generality, we do not formalize the computation language, but instead assume that the function $\s{run}:\s{Thread}\to\s{Operation}$ executes the thread $t$ and returns $op$, which includes a continuation for the rest of the thread (analogous to the coinductive definitions used by \citet{bohannon2009reactive}).

We describe each operation and its corresponding transition rule in turn:
\begin{itemize}
  \item $\rn{s-send}$ for $\opsend{ch}{v}{t'}$:
    This rule checks that the process is permitted to output on channel $ch$ (here, $\s{label}:\s{OutputChannel}\to\s{Label}$ returns the security label of each channel). The process becomes stuck if this check fails. Otherwise, it generates the output event $e=\evsend{ch}{v}$ and the new process state $\process{t'}{l}$ using the continuation $t'$ returned from $\s{run}$.
  \item $\rn{s-read}$ for $\opread{k}{f}$:
    This rule reads the labeled value sequence $\sigma(k)$ from the store; uses the projection operation $\proj{\hole}{l}$ defined in Figure \ref{definition_of_projection} to remove all values not visible to the current label; and passes the last entry in this list (i.e. the most recent visible write) to the read continuation $f$.
    
    Note that $\proj{\sigma(k)}{l}$ may be the empty sequence $\epsilon$, either because key $k$ was never written, or because no such writes are visible to the current process; in this case, $\s{last}(\epsilon)$ returns $\bot$, which is passed to $f$.
    
    This rule (and the following four below) generates a dummy event $\evnoop$, since it does not have any externally visible behavior.

  \item $\rn{s-write}$ for $\opwrite{k}{v}{t'}$:
    In a conventional data store, a new write would overwrite the previous value at that key.
    In contrast, our faceted store semantics must ensure that a low-label process unable to see the new write will still read an older write.
    Hence we represent $\sigma(k)$ as a sequence of labeled values.
    At a new write $(v,l)$ of value $v$ at label $l$, we can garbage-collect or remove all older writes $(v',l')$ in this sequence $\sigma(k)$ that are no longer visible, namely those where $l\sqsubseteq l'$, since any process that could read $(v',l')$ can also read the more recent write $(v,l)$.
    The following function performs this garbage collection, and then appends the new labeled value:
    \begin{center}$\begin{array}{@{}l@{}}
      \s{write} : \s{LabeledValueSeq}\times\s{Value}\times\s{Label}\to
      \\
      \phantom{\s{write}{{}:{}}} \s{LabeledValueSeq}
      \\
      \s{write}(S,v,l) = (S\setminus\{(v',l')\in S~|~l\sqsubseteq l'\})\append(v,l)
    \end{array}$\end{center}
    The symbol $\append{}{}$ denotes sequence concatenation.
  \item $\rn{p-fork}$ for $\opfork{t'}{t''}$:
    This rule forks a new thread $t''$, where $t'$ is the continuation of the original thread. Both threads inherit the security label of the original process.
  \item $\rn{s-raise-label}$ for $\opraiselabel{l'}{t'}$:
    This rule simply raises the label of the current process to  a higher label $l'$, which, for example, permits the process to read more secret data.
  \item
    Finally, $\rn{s-skip}$ allows a state to perform stuttering $\evnoop$ steps at any time, as a technical device to facilitate the non-interference proof below.
\end{itemize}
Note that no rule is needed for the \opstuck{} operation. Instead we just leave the stopped processes in the process multiset for simplicity.

\section{Termination Sensitive Non-Interference}

We use the notation $\proj{\hole}{l}$ on various domains to remove any information that is not visible to an observer at level $l$ (see Figure \ref{definition_of_projection}).
For example: $\proj{S}{l}$ contains only values with labels visible to $l$, and $\proj{ps}{l}$ contains only processes with labels visible to $l$.
An event $e$ \textit{is visible to} $l$ if it starts a process visible to $l$, or if it outputs on a channel visible to $l$; otherwise, we say $\proj{e}{l}=\evnoop$.
We write $\lequiv{l}{\hole_1}{\hole_2}$ to denote that items appear equivalent to an observer at level $l$, i.e., $\proj{\hole_1}{l}=\proj{\hole_2}{l}$.

Our proof is based on the projection lemma below, which relates the execution of the full system $\Sigma$ and the portion $\proj{\Sigma}{l}$ visible at level $l$. Every step of $\Sigma$ has a corresponding step in $\proj{\Sigma}{l}$ (part 1) and vice versa (part 2).

\noindent
\textbf{Lemma (Projection).}
\\
\textbf{Part 1.}
If
\begin{center}
  $\sstep{e_1}{\Sigma}{\Sigma_1'}$
\end{center}
then for some $\Sigma_2'$ and $e_2$,
\begin{center}
  $\sstep{e_2}{\proj{\Sigma}{l}}{\Sigma_2'}$
\qquad
  $\lequiv{l}{\Sigma_1'}{\Sigma_2'}$
\qquad
  $\lequiv{l}{e_1'}{e_2'}$
\end{center}
\textbf{Part 2.}
If
\begin{center}
  $\sstep{e_1}{\proj{\Sigma}{l}}{\Sigma_1}$
\end{center}
then for some $\Sigma_2'$ and $e_2$,
\begin{center}
  $\sstep{e_2}{\Sigma}{\Sigma_2}$
\qquad
  $\lequiv{l}{\Sigma_1}{\Sigma_2}$
\qquad
  $\lequiv{l}{e_1}{e_2}$
\end{center}
\textit{Proof.}
See appendix.

Based on this lemma, our proof of single-step and multi-step termination sensitive non-interference is straightforward.\footnote{
  When applied to a reactive system as we have here, this notion of termination-sensitive non-interference is often known as progress-sensitive non-interference.
}

\noindent
\textbf{Theorem (Single step Termination Sensitive Non-Interference).}
\\
If
\begin{center}
  $\lequiv{l}{\Sigma_1}{\Sigma_2}$
\qquad
  $\sstep{e_1}{\Sigma_1}{\Sigma_1'}$
\end{center}
then for some $\Sigma_2'$ and $e_2$,~~
\begin{center}
  $\sstep{e_2}{\Sigma_2}{\Sigma_2'}$
\qquad
  $\lequiv{l}{\Sigma_1'}{\Sigma_2'}$
\qquad
  $\lequiv{l}{e_1}{e_2}$
\end{center}
\textit{Proof.}
By Projection (Part 1):
\begin{center}
  $\sstep{e_3}{\proj{\Sigma_1}{l}}{\Sigma_3'}$
\qquad
  $\lequiv{l}{\Sigma_1'}{\Sigma_3'}$
\qquad
  $\lequiv{l}{e_1}{e_3}$
\end{center}
for some $\Sigma_3'$ and $e_3$. The $l$-equivalence assumption implies that $\sstep{e_3}{\proj{\Sigma_2}{l}}{\Sigma_3'}$. By Projection (Part 2):
\begin{center}
  $\sstep{e_2}{\Sigma_2}{\Sigma_2'}$
\qquad
  $\lequiv{l}{\Sigma_3'}{\Sigma_2'}$
\qquad
  $\lequiv{l}{e_3}{e_2}$
\end{center}
for some $\Sigma_2'$ and $e_2$. Therefore, by transitivity of $\lequiv{l}{}{}$, we have $\lequiv{l}{\Sigma_1'}{\Sigma_2'}$ and $\lequiv{l}{e_1}{e_2}$, as required.

\noindent
\\
\textbf{Corollary (Termination Sensitive Non-Interference).}
\\
If
\begin{center}
  $\lequiv{l}{\Sigma_1}{\Sigma_2}$
\qquad
  $\sstepstar{es_1}{\Sigma_1}{\Sigma_1'}$
\end{center}
then for some $\Sigma_2'$ and $e_2$,
\begin{center}
  $\sstepstar{es_2}{\Sigma_2}{\Sigma_2'}$
\qquad
  $\lequiv{l}{\Sigma_1'}{\Sigma_2'}$
\qquad
  $\lequiv{l}{es_1}{es_2}$.
\end{center}
\textit{Proof.}
By induction on the derivation of $\sstepstar{es_1}{\Sigma_1}{\Sigma_1'}$.

Our Coq formalization of these semantics and proofs is available at \url{https://github.com/anonymous-pldi2018-1/faceted-tsni-coq}.

Our TSNI result states that the set of observable outputs of the 
system under \emph{all possible schedules} does not depend on inputs 
that are not visible to the observer. This result does not prevent a 
malicious scheduler from leaking secrets by prioritizing certain 
schedules, e.g., by scheduling low-security processes based on 
high-security secrets. However, as mentioned earlier, we assume that 
the scheduler is not adversarial in this manner. Prior work 
\cite{Stefan-etal-12-ICFP} has addressed this problem by assuming a 
round-robin scheduler, but this assumption is not realistic for 
serverless computing.  
 
%Note that if a deterministic implementation implements a refinement 
%$(\Longrightarrow)\subset(\longrightarrow)$ of the transition 
%relation, then our TSNI result may not apply to the deterministic 
%transition relation. For example, $\Longrightarrow$ could schedule 
%low-security processes based on high-security secrets.  However, as 
%mentioned earlier, we assume that the scheduler is not adversarial in 
%this manner. Prior work \cite{Stefan-etal-12-ICFP} has addressed this 
%problem by assuming a round-robin scheduler, but this assumption is 
%not realistic for serverless computing.

%% file: evaluation.tex
\section{Implementation and evaluation}\label{sec:evaluation}

\subsection{Implementation}\label{sec:implementation}

In order to evaluate our proposed security architecture and IFC model, 
we developed an open-source prototype implementation of 
\system~\cite{Trapeze-github}.  The implementation is portable and 
currently runs on the two most popular serverless platforms---AWS 
Lambda~\cite{AWSLambda} and OpenWhisk~\cite{OpenWhisk}.  It consists 
of three components: the sandbox, the security shim, and the 
authentication service.  

\mypara{Sandbox} The \system sandbox encapsulates application 
code, redirecting all its inputs and outputs through the security shim 
(Figure~\ref{fig:arch}).  The exact sandboxing technology depends on 
the programming language used.  We currently support serverless 
functions written in JavaScript for the Node.js runtime, which is 
one of the most common types of serverless functions on both AWS and 
IBM Cloud Functions, IBM's public OpenWhisk service. We encapsulate such functions using the VM2 JavaScript 
sandbox~\cite{VanCutsem-13,VM2}.  Many serverless functions rely on 
the ability to invoke external programs.  We encapsulate such external 
executables using a \src{ptrace}-based sandbox~\cite{gg17}, which 
restricts program's I/O activity to a temporary local directory that 
gets purged on every serverless function invocation.

\mypara{Security shim}  The security shim monitors all inputs and 
outputs of a function and enforces IFC rules.
%It does all this transparently to the application, which is unaware 
%of security labels and faceted store.
The shim consists of multiple adapter modules, one for each supported 
input and output interface.  There are three groups of adapters: (1) 
data store adapters, (2) function call adapters, (3) external channel 
adapters. 

A data store adapter implements faceted store semantics on top of a 
conventional cloud data store.  \system currently supports a single 
type of data store---a faceted key-value store implemented on top of a 
relational database.  The key-value store implements a standard 
dictionary with the following operations: \src{put(key, value)}, 
\src{get(key)}, \src{del(key)}, and \src{keys()} (returns all  keys in the 
store).
%, and a convenience method that returns all stored key-value 
%pairs.
The store is backed by a relational database table with 3 
columns, for the key, value, and label.  The table contains an entry 
for each facet of each value in the store.  We used MySQL server 
available in AWS through Amazon Relational 
Database Service.  The security shim passes an additional 
parameter to every operation---the security label.  The \src{get} 
operation performs an SQL query that returns all entries that match 
the given key, and whose label is less than or equal to the given 
label. The \src{del} operation deletes all facets with labels greater 
than or equal to the given label.
%(all rows in the table with the given 
%key, and a label greater than or equal to the given label).
The \src{put} operation deletes the same elements a \src{del} 
does, and then inserts the given key-value pair, with the 
given label.  In addition to the faceted version, we also provide a 
conventional insecure key-value store implementation as a baseline for 
performance evaluation.

Function call adapters support different ways to invoke serverless 
functions, making sure that the callee inherits the caller's  
label, as required by our IFC model.  We support two 
invocation mechanisms: AWS Step Functions, which  
run a workflow with multiple serverless functions, 
controlled by a finite automaton, and Amazon Kinesis, which supports 
asynchronous communication via real-time event streams.

External channel adapters enable secure communication across the cloud 
boundary.
%while blocking access to external channels whose labels are not 
%greater than or equal than the function's label.
The supported types 
of channels are (1) user-initiated HTTP sessions, (2) email 
communication via the Nodemailer module\cite{nodemailer}, and (3) 
connections to external S3 buckets (used to upload large data objects 
that do not fit in HTTP requests).  HTTP sessions obtain their 
security labels from the authentication service (see below).  The 
Nodemailer adapter uses the user database (see below) to map an email 
address to a user security label.  The S3 adapter inherits the label 
of the user who provides login credentials for the S3 bucket.

\mypara{Authentication service}  The authentication service is 
responsible for associating a correct security label with every 
external HTTP session.  It is implemented on top of a user database 
that stores credentials, email addresses, and security labels of all 
users in the system.

The entire \system framework consists of 1174 lines of JavaScript code, 
including 
%XXX lines in the sandbox (excluding VM2 code), % There isn't any sandbox code that is not also shim code.
649 lines in the the AWS and OpenWhisk shim modules, 484 lines in the key-value store, and 41 lines in the authentication service.

\subsection{Evaluation questions}

Our evaluation aims to answer the following questions:
\begin{enumerate}
    \item \emph{Security:} Can \system enforce information 
        security in real-world serverless applications?  In 
        particular, can confidentiality requirements of such 
        applications be captured in a 
        security policy consisting of a security lattice and trusted 
        declassifiers?  Can \system enforce the policy in the presence 
        of buggy or malicious code?
        
    \item \emph{Transparency:} Can \system secure existing serverless 
        applications with minimal modifications?

    \item \emph{Performance:} Can \system achieve the first two goals 
        with low performance overhead?
\end{enumerate}

\subsection{Case studies}

To answer the above questions, we carried out three case studies where 
we used \system to add a security layer to existing serverless 
applications.  We outline each of the case studies below.

\mypara{Case study 1: \syshello}
\syshello~is a project from the serverless team at 
Nordstrom. The goal was to produce a purely 
serverless, back-end for an e-commerce web 
site. It has since been 
open-sourced~\cite{HelloRetail} and won the architecture competition 
award at serverlessConf Austin'17\cite{HelloRetail-award}.
%, and is consistently considered one of the largest publicly 
%available examples of serverless applications. % Would really like to 
%justify this claim with a reference, but can't find one.

\begin{figure}
    \center
    \includegraphics[width=\linewidth]{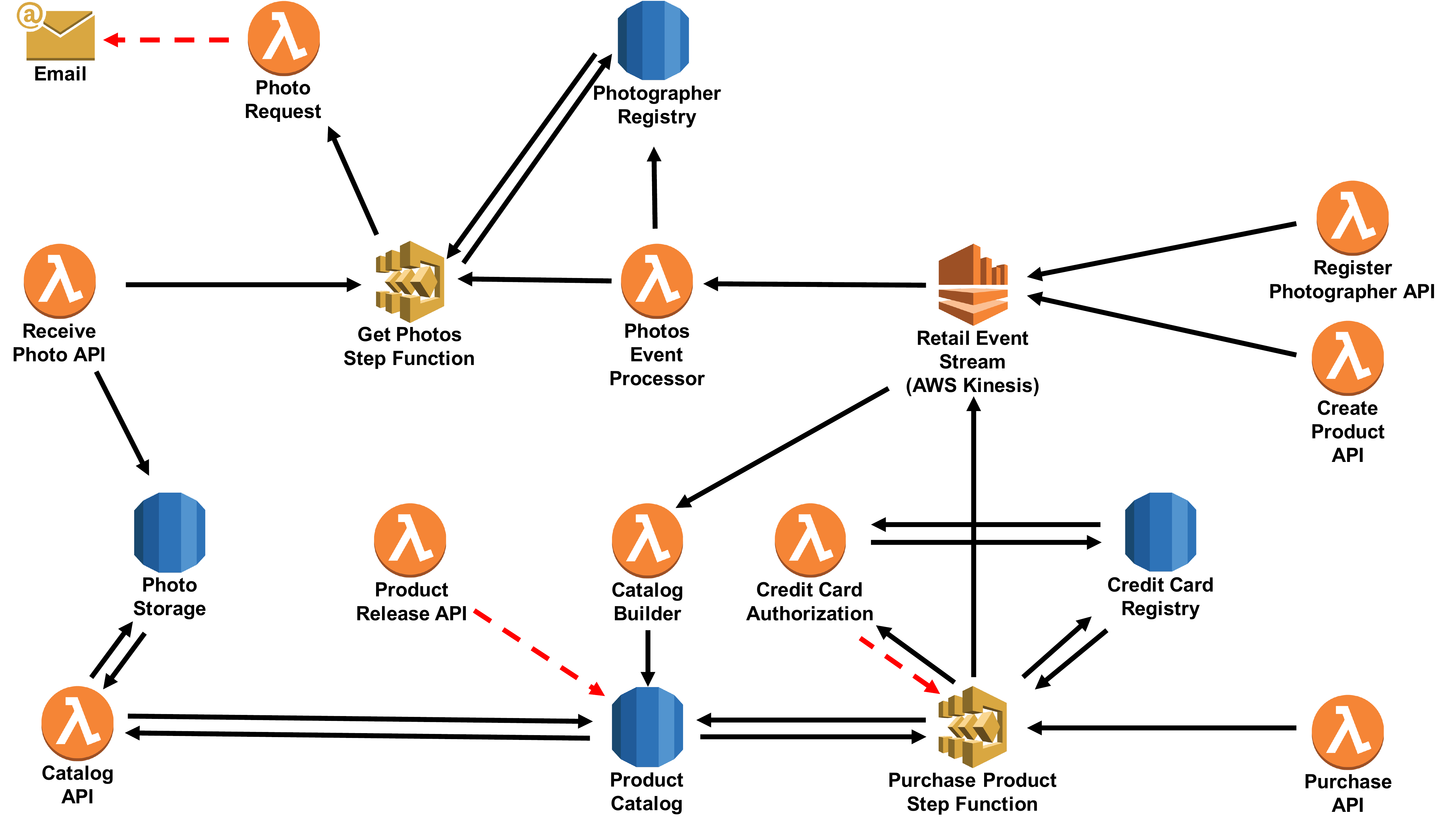}
    \caption{The architecture of the \syshello~project.  
    Circles labeled $\lambda$ represent the main functions of 
    this application.  Functions whose names end with `API' can be directly 
    invoked via client HTTP requests.  Arrows show interactions 
    between different components.  Red dashed 
    arrows indicate interactions that carry declassified 
    data.}\label{fig:hello-retail}
\end{figure}

We made several changes to \syshello~before applying 
\system to it.  First, we replaced DynamoDB and S3 databases in 
\syshello~with with calls to \system's key-value store.  Second, we replaced 
calls to the Twilio SMS messaging service, which is currently not 
supported by \system, with e-mail communication.  Third, we extended 
the \syshello~project with a product purchase subsystem, which manages 
online orders and credit card payments. 
The resulting system consists of 21 serverless functions.  
Figure~\ref{fig:hello-retail} shows the high-level architecture of the 
system.  

%Note that the `Get Photos' and `Purchase Product' step 
%functions are state machines with 8 and 5 serverless functions 
%respectively.

\begin{figure}
\begin{subfigure}[b]{0.5\textwidth}
    \centering
    \includegraphics[width=0.9\textwidth]{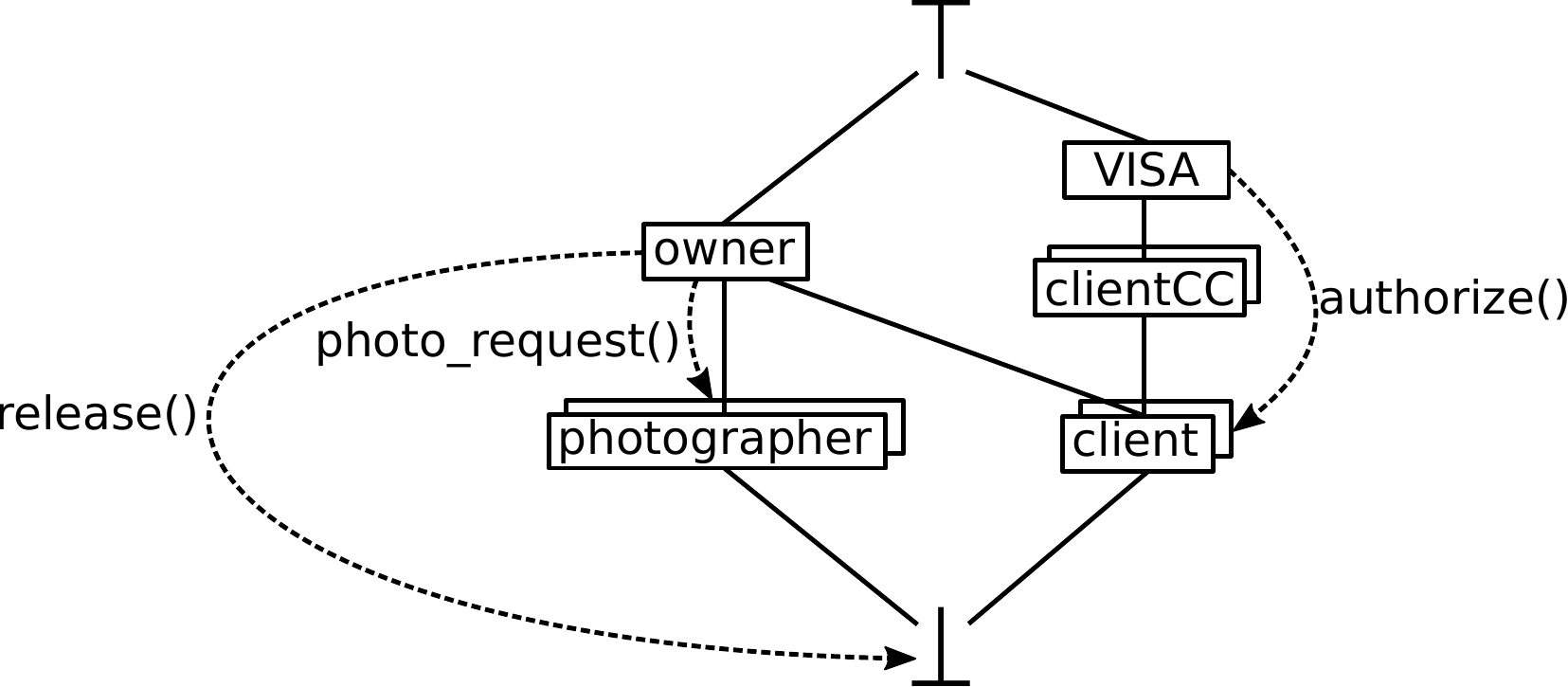}
    \caption{\syshello.}\label{fig:nordstrom_lattice}
\end{subfigure}
~
\begin{subfigure}[b]{0.2\textwidth}
    \centering
    \includegraphics[width=0.9\textwidth]{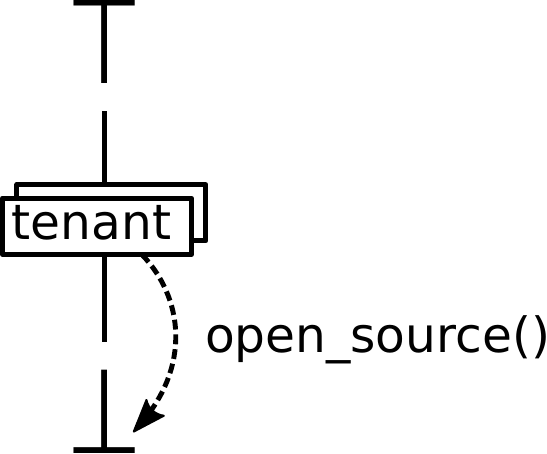}
    \caption{\sysgg}\label{fig:gg_lattice}
\end{subfigure}
~
\begin{subfigure}[b]{0.2\textwidth}
    \centering
    \includegraphics[width=0.35\textwidth]{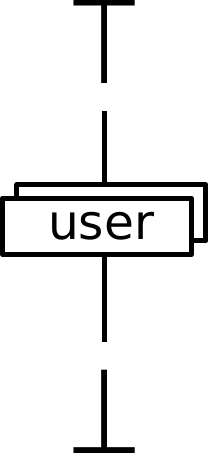}
    \caption{Feature Extraction.}\label{fig:image_lattice}
\end{subfigure}
\caption{Security lattice and declassifiers in the different case studies.}
\end{figure}

The system serves several types of users: (1) the store owner, who 
manages the online catalog and processes orders, (2) 
photographers, who upload product images to the catalog, (3) 
customers, who navigate the catalog and place orders, and (4) 
the VISA credit card authority, which authorizes card payments on 
behalf of customers.  The security lattice 
(Figure~\ref{fig:nordstrom_lattice}) consists of labels, matching 
these user categories.  Solid lines in the diagram show the partial 
order of security labels; dashed arrows show 
declassifiers, with a declassifier $\langle h, l, D \rangle$ 
represented by an arrow from label $h$ to $l$.  
Table~\ref{tab:nordstrom_labels} summarizes the security labels in 
this case study.

\begin{table}
    \small
    \centering
    \begin{tabular}{|l|p{0.7\linewidth}|}
        \hline
        \textbf{label}   & \textbf{description} \\
        \hline\hline
        $owner$          & Sensitive information managed by the store owner, including product catalog and photographers' email addresses.
                           Items in the catalog are only visible to the owner until they are released to the public via the \src{release} declassifier.\\
        $client_j$       & Labels online purchases of a specific customer.
                           This information is visible to the customer and the store owner (since $client_j \sqsubset owner$)\\
        $clientCC_j$     & Customer credit card information that can only be released to a credit card authority\\
        $VISA$           & Labels the external communication channel to the VISA credit card authority\\
        $photographer_i$ & Product photos uploaded by a photographer \\
        \hline
    \end{tabular}
    \caption{Security classes in the \syshello~case study.}\label{tab:nordstrom_labels}
\end{table}

The following scenario illustrates the
flow of sensitive information through the \syshello~system.  
Every step in the scenario is annotated with security label(s) of data 
involved in this step.

\noindent 1. [$owner$] The owner creates a new product description in 
        the catalog.\\
\noindent 2. [$photographer_i$] The owner sends an email to one of the 
        photographers requesting a picture of the product.  The 
        request includes information from the product description,
        declassified with the $\langle owner, photographer_i, 
        photo\_request()\rangle$ declassifier, which implements a 
        trusted user interface that request owner's confirmation of 
        the declassification.\\
\noindent 3. [$photographer_i$] The photographer uploads a product 
        image to the catalog.\\
\noindent 4. [$\bot$] Once the owner is ready to make the product 
        publicly available in the online catalog, she declassifies it 
        using the $\langle owner, \bot, release()\rangle$ 
        declassifier.\\
\noindent 5. [$client_j/clientCC_j$] A client orders a product from the 
        catalog.  Order information, labeled $client_j$ is 
        visible to the the client as well as the owner, since 
        $client_j \sqsubset owner$.  Client's credit card details are 
        labeled $clientCC_j$, and are hidden from the owner 
        ($clientCC_j \not\sqsubseteq owner$).\\
\noindent 6. [VISA] Before the order is finalized, credit card 
        information is sent to VISA for payment authorization through 
        an external channel labeled $VISA$.\\
\noindent 7. [$client_j$] The response received through this channel 
        consists of one bit of information indicating success or 
        failure, which gets declassified by the $\langle VISA, 
        client_j, authorize()\rangle$ declassifier, making the outcome 
        of the request visible to the client and the owner.

%Once secured, the faceted store semantics allowed us to trivially 
%implement a `Product Release' subsystem that leverages the fact that 
%the system hides privileged information from unprivileged users. An 
%added product has an `owner' security label, and so are invisible to 
%clients. By adding a declassifier that changes the label of a product 
%from `owner' to `public' we were able to add the required 
%functionality without making any changes to the storage schemas or 
%the front-end.

%The security policy enforced in the Hello, Retail! case is 
%illustrated in Figure~\ref{fig:hello-retail-lattice}. This lattice 
%encodes the requirements that photographer private information not be 
%accessible to cients, and vice-versa, as well as the requirement that 
%stored credit card information only be accessible to the credit card 
%company.

%We ran the Hello, Retail! use cases with several scenarios \TODO{list 
%scenarios.}

%\begin{figure}
%    \center
%    \includegraphics[width=0.9\linewidth]{img/hello-retail-lattice}
%    \caption{The security policy enforced in the Hello, Retail!  
%    example. Solid edges show the partial order between the security 
%    labels. Dashed edges show the declassifiers present in the 
%    system.}\label{fig:hello-retail-lattice}
%\end{figure}

\mypara{Case study 2: \sysgg}
\sysgg~\cite{gg17,gggithub} is a system for running parallel software 
workflows, such as software compilation and video processing, on 
serverless platforms. In \sysgg, each unit of work, or \emph{thunk}, 
specifies both the executable to run and all its data dependencies.  
The workflow is synthesized as a direct acyclic graph (DAG) of thunks, 
and is recursively executed on a serverless platform by \sysgg's 
execution engine.  \sysgg identifies each dependency in a 
content-addressed way, using a key-value store as the storage backend.  
\sysgg consists of a single serverless function, which internally runs 
arbitrary user-provided executables.  Each invocation of the function 
executes exactly one thunk by fetching the dependencies from the 
object store, executing the thunk, and storing the output back into 
the object store. 

We use a parallel build framework implemented on top of 
\sysgg~\cite{gg17} as a concrete use case.  The framework extracts a 
workflow DAG, where every thunk corresponds to an invocation of a 
build tool (e.g., a compiler or a linker), from the project makefile.
%, and executes each thunk via a separate serverless call.

The original \sysgg implementation is single-tenant, with every 
authenticated user having access to all sources and binaries in the 
system.  We use \system to introduce a secure multi-tenant mode to 
\sysgg.  In this mode, tenants only have access to their own source 
and compiled code.  A thunk running on behalf of the tenant taints all 
of its outputs with the tenant's label.  A tenant may release some of 
their sources or compiled binaries to the public, making them 
available to all other tenants.  This is reflected in the security 
lattice in Figure~\ref{fig:gg_lattice} with mutually incomparable 
tenant labels and a $\langle user_i, \bot, open\_source() \rangle$ 
declassifier.

Prior to adding the multi-tenant mode to \sysgg, we ported parts of it 
that were written in Python to JavaScript, as well as modified it to 
use our key-value store. 

\mypara{Case study 3: Image feature extraction}
This serverless application gives its users access to Amazon's AWS 
Rekognition image analysis service~\cite{AWSRekognition}.  It is based on the `Fetch File and Store 
in S3' and `Analyse Image from S3' examples from the Serverless 
Examples collection\cite{Serverless-examples}.

The application consists of the upload function that 
takes an image URL, fetches the image and stores it in the key-value 
store, and the feature extraction function that uses AWS Rekognition 
to extract features from the image and send them to the user.

We use \system to add a security layer to this example, enforcing that 
every user can only access information extracted from images they own.  
This policy is expressed in the simple security lattice in 
Figure~\ref{fig:image_lattice}.

\subsection{Security}

We employ \system to protect sensitive data in the three case studies.  
The original implementation of these applications either offered 
coarse-grained protection, giving every authenticated user access to 
all data in the system, or implemented ad hoc security policies 
embedded in application code.  For example, the \syshello~system by 
design only exposes credit card details to the credit card authority.  
Such protection relies on checks scattered around the application code 
and is easy to get wrong.  Besides, it can be bypassed by any exploit 
that subverts the application logic.

\system captures the security requirements of each of the three case 
studies in a security policy consisting of a security lattice and 
declassifiers, shown in Figures~\ref{fig:nordstrom_lattice}, 
\ref{fig:gg_lattice}, \ref{fig:image_lattice}.  Our policies are 
simple and concise, consisting of only several classes of labels and 
few declassifiers.  The policies are decoupled from the application 
logic and its software architecture.  For instance, adding new 
functions to the application, changing its control flow, or even 
refactoring the
database schema, do not affect the security policy.

Furthermore, \system is immune to malicious or compromised application 
logic.  We simulated code injection attacks in our case studies by 
replacing some of the original functions with malicious functions that 
attempt to leak secrets to unauthorized users, similar to examples in 
Figures~\ref{fig:exploit3} and~\ref{fig:exploit2}.  As expected, these 
simulated attacks failed when running the application with \system.

\subsection{Transparency}

As discussed in Section~\ref{sec:design}, \system enforces TSNI at the
cost of reduced transparency, which may require the developer to refactor 
the application to work with \system.

Table~\ref{tab:locs} measures the loss of transparency in our case 
studies by reporting the size of changes to application code in terms 
of lines of code (LoCs) required to adapt the application to work with 
\system.  This does not take into account the changes needed to port 
the application to use our key-value store as well as other 
compatibility changes that are not fundamental to the \system 
architecture and can be made unnecessary with some additional 
engineering effort.

\begin{table}
    \small
    \begin{tabular}{|l|c|c|c|}
        \hline
        \textbf{case study} & \textbf{app code} & \textbf{modif. code} & \textbf{declassifiers} \\
        \hline \hline
        \syshello          & 1,300 & 3 & 104 \\
        \sysgg             & 8,800 & 2 &  94 \\
        Feature Extraction &    95 & 0 &   0 \\
        \hline
    \end{tabular}
    \caption{Size of the case studies in lines of code (LoCs): (1) 
    LoCs comprising the application, (2) lines added or modified to 
    adapt the application to work with \system, (3) LoCs in trusted 
    declassifiers.}\label{tab:locs}
\end{table}

None of the case studies required splitting a function into multiple 
functions.  \syshello~required calling \textsf{raiseLabel} twice, once 
in order to upgrade the label of the purchase placement function from 
$client_j$ to $clientCC_j$ before saving the customer's credit card 
details in the data store, and once to upgrade the label of the 
payment authorization function from $client_j$ to $VISA$, in order to 
read credit card details from the store and send them to the credit 
card authority.  In addition, the \src{gg} case study required a minor 
change due to a technicality: the existing code was not fully 
compatible with the VM2 sandbox.

These results indicate that in practice the loss of transparency is 
not an issue in serverless applications.  This is due to the common 
software design practices in the serverless world, where each function 
is assigned a single small task and only accesses the values that are 
related to this task and are therefore likely to have compatible 
security labels.

The last column of the table reports the total size of declassifiers 
used in each case study.  This number characterizes the amount of 
trusted application code in each example.

\subsection{Performance}

We measure the overhead of \system by running case studies 1 and 3 on 
the AWS Lambda serverless platform.  We run case study 2 on the 
OpenWhisk platform in the IBM Cloud (formerly, IBM Bluemix), since AWS
Lambda does not support \texttt{ptrace}, which we use to sandbox 
binary executables in \sysgg (Section~\ref{sec:implementation}).

Table~\ref{tab:perf} summarizes the runtime overhead of \system.  
Since the overhead may depend on the exact workload, we constructed 
several typical workflows for each case study.  We test the 
\syshello~case study on two workloads: the `Build\&Browse' workload, 
that simulates the construction and browsing of the product catalog, 
and the `Update\&Purchase' workload, that simulates updates to the 
catalog followed by a series of online purchases.  We test \sysgg by 
using it to compile four open source software packages: mosh, git, 
vim, and openssh.  Finally, we consider two scenarios for the Image 
Feature Extraction: the `Image Upload' scenario, that uploads a single 
image to the data store, and the `Feature Extraction' scenario that 
performs feature extraction for a stored image.

\begin{table}
    \small
    \begin{tabular}{|l|c|c|c|c|}
        \hline
        \multirow{2}{*}{\textbf{scenario}} & \multirow{2}{*}{\textbf{$\#\lambda$}} & \multicolumn{2}{|c|}{\textbf{function runtime (ms)}} & \multirow{2}{*}{\textbf{$\Delta$(\%)}} \\
        \cline{3-4}
                                           & & \textbf{insecure} & \textbf{\system} & \\
        \hline \hline
        \multicolumn{5}{|c|}{\syshello} \\
        \hline
        Update\&Purchase      & 435  & 51,246  &  98,563 & 92.33 \\
        Build\&Browse         & 955  & 257,799 & 391,900 & 52.03 \\
        \hline
        \multicolumn{5}{|c|}{\sysgg} \\
        \hline
        mosh    & 111 &  681,173  & 654,448   & -3.92  \\
        git     & 917 & 2,602,500 & 2,660,721 &  2.24  \\
        vim     & 234 & 1,242,873 & 1,338,128 &  7.66  \\
        openssh & 654 & 1,626,223 & 1,649,139 &  1.41   \\
        \hline
        \multicolumn{5}{|c|}{\sysimageext} \\
        \hline
        Image Upload       & 1 & 475 &  525 & 10.4   \\
        Feature Extraction & 1 & 1,882 & 2,114 & 12.3  \\
        \hline
    \end{tabular}
    \caption{\system performance.}
    %\KALEV{A comment about the Catalog Build \& Browsing scenario -- since we have additional lambdas running in the secure version (the product release subsystem), the result is kind of skewed. If we compare the per-lambda overhead, then it drops from 68.18\% to 42.58\%.}
    \label{tab:perf}
\end{table}

For each workload, we report the total number of serverless function 
calls (the $\#\lambda$ column in Table~\ref{tab:perf}).  The 
`function runtime' section of the table reports the total runtime of 
all functions in the scenario with and without protection and the 
relative slowdown introduced by protection, averaged across 10 runs.  These runtimes do not 
include 
%the overhead of remote function calls and the
additional declassifier calls introduced in the protected execution.  
Accounting for declassifiers increased the total runtime by another 
10\% in the Build\&Browse workload, while not making any measurable
impact on other workloads.
In our initial experiments we also measured the total time to execute 
each workload; however we found that, due to the non-deterministic nature 
and varying resource availability in the serverless environment, these times
varied wildly across different runs and did not provide any insights into the performance 
of \system.

%Note that this time can be smaller than the total function runtime due 
%to parallelism in the serverless runtime.

For most workloads, \system adds a modest overhead of up to $12.3\%$ 
to function runtime.  The negative overhead in the \sysgg mosh workload is 
due to the noisy IBM Cloud environment, where the runtime of a function 
varies dramatically across different invocations based on the load on the 
node where the function is scheduled.  The overhead is higher in the \syshello case 
study, up to 92.33\% in the Update\&Purchase scenario.  Further 
benchmarking revealed that the bulk of this overhead is due to the 
startup time of the VM2 sandbox, which adds 100ms on average per
function invocation.  Since all functions in this example have short
runtimes, the startup time  becomes a significant contributor to the total 
runtime. We measured that two thirds of the startup time is spent 
loading libraries used by the application.
%\LEONID{Kalev, do we have data to make a statement along the lines of: 
%``XXX\% of this overhead is due to the startup delay''}\KALEV{Yes. I ran an experiment with a lambda that does nothing but load a bunch of libraries. The insecure lambda runs 1.59ms on average, the secure lambda runs 156.52ms, the secure lambda without the module load runs 50.33ms. The per-delta overhead in the first scenario in \syshello is 111ms.}
In the future, much of this overhead can be eliminated by caching preloaded libraries 
in memory.

Finally, we evaluate the storage overhead of our secure key-value 
store.  Table~\ref{tab:storage} compares database sizes (in kilobytes) 
for an insecure and secure versions of the store for each of the three 
case studies (we report results for one of the workloads in each case 
study, as the relative increase in the database size is independent of 
the workload).  The secure database requires more space, as it stores 
a security label with each value.  This overhead is low (under $2\%$) 
in examples where the database stores large objects, e.g., images or 
source files.  It is more significant (18.34\%) in the \syshello case 
study where individual values stored in the database are small.

\begin{table}
    \small
    \begin{tabular}{|l|c|c|c|}
        \hline
        \textbf{case study} & \textbf{insecure (KB)} & \textbf{\system (KB)} &  \textbf{$\Delta$(\%)} \\
        \hline \hline
        \syshello          &     2,704  &      3,200 &  18.34 \\
        \sysgg             & 15,054,000 & 15,118,496 &  0.43  \\
        Feature Extraction &      8,256 &      8,416 &  1.94  \\
        \hline
    \end{tabular}
    \caption{Storage overhead of \system.}\label{tab:storage}
\end{table}

%% file: related.tex
\section{Related work}\label{sec:related}

The modern technique for using security labels to dynamically monitor information flow was proposed by \citet{Denning-76-CACM} as the Lattice Model.
That work also defines the concept of \textit{implicit flow} of information.
\citet{Austin-etal-17-TOPLAS} describe three techniques for monitoring such flows, namely \textit{Failure Oblivious} and \textit{No Sensitive Upgrade} and \textit{Permissive Upgrade}.
Our work exposes a fourth choice using faceted values specifically to eliminate implicit flows.

\citet{Austin-Flanagan-12-POPL} introduce faceted values as a full language-based enforcement technique: rather than focusing specifically on controlling implicit flows, they combine the faceted value concept with the concept of \textit{multiple executions}, which enables more precise enforcement at the cost of more runtime overhead.

\system enforces a strong form of non-interference, 
TSNI.  Most existing IFC 
systems support only termination-insensitive non-interference (TINI).  The few 
previous systems that enforce TSNI are either too restrictive or too 
costly for most practical applications.  The \emph{multilevel 
security} (MLS) model~\cite{Bell-LaPadula-73} achieves TSNI by 
statically partitioning all the code and data in the system into 
security compartments.  This model, designed primarily for 
military-grade systems, is too restrictive for most applications and 
requires complete re-design of existing software.  Smith and 
Volpano~\cite{Smith-Volpano-98} present a security type system that 
enforces TSNI by imposing a harsh restriction that loop conditions may 
not depend on secret data.  Heintze and Riecke~\cite{Heintze-98} propose a secure typed 
Lambda calculus called SLam.  While SLam only 
enforces TINI, Sabelfeld and Sands~\cite{Sabelfeld-Sands-01} point out 
that a version of SLam with lazy evaluation semantics would be 
termination-sensitive.  This theoretical result has 
limited practical implications, as, with the exception of Haskell, 
none of today's major programming languages use lazy evaluation.

Indeed, \citet{Stefan-etal-12-ICFP} implemented a Haskell library called \textit{LIO}, which guarantees TSNI by requiring programmers to decompose their programs into separate threads with floating labels.
These are analogous to the processes in our formalism.
However, LIO uses the No Sensitive Upgrade rule to prevent implicit flows, while \system uses faceted values instead. They also assume a round-robin scheduler, which would be inappropriate for our application to serverless computing.

\emph{Secure multi-execution}~\cite{Devriese-Piessens-10-SSP} achieves 
TSNI by running multiple independent copies of the program, one for 
each security class.  This technique introduces CPU and memory 
overhead proportional to the number of security classes.  While 
acceptable in systems with few security classes (e.g., Devriese and 
Piessens~\cite{Devriese-Piessens-10-SSP} consider only two classes, 
secret and non-secret), this becomes impractical in 
cloud-scale systems with potentially millions of mutually untrusting 
users.
Faceted execution has the potential to mitigate this drawback, and \citet{bielova2016spot} have proposed a theoretical approach for extending the faceted execution model to enforce TSNI.
%Our approach does not use multiple executions, but in exchange, we do not offer the same transparency guarantees.

The Asbestos OS~\cite{Efstathopoulos:2005} applies dynamic IFC at the 
granularity of an OS process, similar to 
how \system operates at the granularity of a serverless function.  
Asbestos associates a static security label with each process; however 
this label only serves as an upper bound on the label of data 
the process can access.  Process's effective label changes 
dynamically, which enables the implicit termination channel.  

To the best of our knowledge, \system is the first system to apply IFC 
to serverless applications.  Several researchers advocate the use of 
IFC in the broader context of secure cloud 
computing~\cite{Bacon-14,Pasquier-16}; however we are not aware of a 
practical implementation of these ideas.

%% file: conclusion.tex
\section{Conclusion}

The advent of serverless computing provides the opportunity to rebuild
our cloud computing infrastructure based on a rigorous foundation for
information flow security. We present a novel and promising approach
for dynamic IFC in serverless systems. This approach combines (1) a 
sandbox and security shim that monitors all I/O operations of
each serverless function invocation; (2) static security labels for 
each serverless function invocation; and (3) dynamic faceted labeling 
of data in the persistent store.

This combination of ideas provides the strong security guarantee of
TSNI, which is necessary in serverless settings to avoid
high-bandwidth termination channel leaks via multiple concurrent
requests.

Our \system implementation of this approach is lightweight, requiring
no new programming languages, compilers, or virtual machines. The
three case studies show that \system can enforce important IFC
properties with low space and time overheads.

%of less than 20% in all cases.

We believe \system represents a promising approach for deploying
serverless systems with rigorous security guarantees that help prevent
costly information leaks arising from buggy or mis-configured
application code or from code-injection attacks.

%% file: acks.tex
\begin{acks}                            %% acks environment is optional
                                        %% contents suppressed with 'anonymous'
  %% Commands \grantsponsor{<sponsorID>}{<name>}{<url>} and
  %% \grantnum[<url>]{<sponsorID>}{<number>} should be used to
  %% acknowledge financial support and will be used by metadata
  %% extraction tools.
  % This material is based upon work supported by the
  % \grantsponsor{GS100000001}{National Science
  %   Foundation}{http://dx.doi.org/10.13039/100000001} under Grant
  % No.~\grantnum{GS100000001}{nnnnnnn} and Grant
  % No.~\grantnum{GS100000001}{mmmmmmm}.  Any opinions, findings, and
  % conclusions or recommendations expressed in this material are those
  % of the author and do not necessarily reflect the views of the
  % National Science Foundation.
\end{acks}

%% file: appendix_proof.tex
\section{Summary of auxiliary semantic details}

$\begin{array}{@{}l}
  \text{$\epsilon$ is the empty sequence and $\append{}{}$ is concatenation.}
  \\
  \s{label} : \s{OutputChannel}\to\s{Label}
  \\
  \s{run} : \s{Thread}\to\s{Operation}
  \\
  \s{write} : \s{LabeledValueSeq}\times\s{Value}\times\s{Label}\to
  \\
  \phantom{\s{write}{{}:{}}} \s{LabeledValueSeq}
  \\
  \s{write}(S,v,l) = (S\setminus\{(v',l')\in S~|~l\sqsubseteq l'\})\append(v,l)
  \\
  \text{$\lequiv{l}{\hole_1}{\hole_2}$ means $\proj{\hole_1}{l}=\proj{\hole_2}{l}$}
\end{array}$

\section{Proof details}

\textbf{Lemma (Invisibility).}
\\
If
\begin{center}
  $l'\not\sqsubseteq l$
\\
  $\sstep{e}{(\sigma,ps\uplus\{\process{t}{l'}\})}{(\sigma',ps\uplus ps')}$
\end{center}
then
\begin{align*}
  \proj{\sigma'}{l}&=\proj{\sigma}{l}
\\
  \proj{e}{l}&=\evnoop
\\
  \proj{ps'}{l}&=\{\}
\end{align*}
\textit{Proof.}
Omitted.

\noindent
\textbf{Lemma (Projection 1).}
\\
\startproof
  (1) & If $\sstep{e_1}{\Sigma}{\Sigma_1'}$,
\\
      & then $\exists\Sigma_2'.\exists e_2.~\sstep{e_2}{\proj{\Sigma}{l}}{\Sigma_2'}$
        and $\lequiv{l}{\Sigma_1'}{\Sigma_2'}$
        and $\lequiv{l}{e_1'}{e_2'}$.
\finishproof
\\
\textit{Proof.}
\\
\newcommand{\z}{$\quad$}
\startproof
        & Let $(\sigma, ps) = \Sigma$.
\\
        & Let $(\sigma_1', ps_1') = \Sigma_1'$.
\\
        & Proceed by cases (ie inversion) on (1).
\\
        & Case: $\rn{s-start}$. Let $t,l'$ be such that:
\\
        & \z $e_1=\evstart{\process{t}{l'}}$; and
\\
        & \z $\sigma_1'=\sigma$; and
\\
        & \z $ps_1'=ps+\process{t}{l'}$.
\\
        & \z Proceed by cases.
\\
        & \z Case: where $l'\sqsubseteq l$.
\\
        & \z \z Pick $\Sigma_2' = {(\proj{\sigma}{l},\proj{ps}{l}+\process{t}{l'})}$.
\\
        & \z \z Pick $e_2' = {\evstart{\process{t}{l'}}}$.
\\
        & \z \z QED
        & by  $\rn{s-start}$.
\\
       & \z Case: where $l'\not\sqsubseteq l$.
\\
        & \z \z Pick $\Sigma_2' = \proj{\Sigma}{l}$ and pick $e_2' = \epsilon$.
\\
        & \z \z QED
        & by  $\rn{s-skip}$.
\\
        & Case: $\rn{s-skip}$. $e_1=\epsilon$ and $\Sigma_1'=\Sigma$.
\\
        & \z Pick $\Sigma_2' = \proj{\Sigma}{l}$ and pick $e_2' = \epsilon$.
\\
        & \z QED
        & by  $\rn{s-skip}$.
\\
        & \multicolumn{2}{@{}l}{Last case: any other rule. Let $t,l',ps_1,ps_2$ be such that:}
\\
        & $ps=ps_1\uplus\{\process{t}{l'}\}$; and
\\
        & $ps_1'=ps_1\uplus ps_2$; and
\\
        & Proceed by cases.
\\
        & Case: where $l'\not\sqsubseteq l$.
\\
        & \z $\proj{\sigma_1'}{l}=\proj{\sigma}{l}$ and $\proj{e_1}{l}=\epsilon$ and $\proj{ps_2}{l}=\{\}$
        & by Invisibility.
\\
        & \z QED
        & by $\rn{s-skip}$.
\\
        & Last case: where $l'\sqsubseteq l$.
\\
        & Resume case analysis on (1).
\\
        & Case: $\rn{s-send}$. Let $ch,v,t'$ be such that:
\\
        & \z $e_1=\evsend{ch}{v}$; and
\\
        & \z $\sigma_1'=\sigma$; and
\\
        & \z $ps_2=\{\process{t'}{l'}\}$; and
\\
        & \z $run(t)=\opsend{ch}{v}{t'}$; and
\\
        & \z $l'\sqsubseteq\s{label}(ch)$.
\\
        & \z Pick $\Sigma_2' = (\proj{\sigma}{l},\proj{ps_1}{l}\uplus\{\process{t'}{l'}\})$.
\\
        & \z Pick $e_2' = {\evsend{ch}{v}}$.
\\
        & \z QED
        & by $\rn{s-step}$.
\\
        & Remaining cases omitted.
\finishproof

\noindent
\textbf{Lemma (Projection 2).}
\\
\startproof
  (1) & If $\sstep{e_1}{\proj{\Sigma}{l}}{\Sigma_1}$,
\\
      & then $\exists \Sigma_2.\exists e_2.~\sstep{e_2}{\Sigma}{\Sigma_2}$
        and $\lequiv{l}{\Sigma_1}{\Sigma_2}$
        and $\lequiv{l}{e_1}{e_2}$.
\finishproof
\\
\textit{Proof.}
\\
\startproof
        & Let $(\sigma,ps)=\Sigma$.
\\
        & Let $(\sigma_1,ps_1)=\Sigma_1$.
\\
        & Proceed by cases (ie inversion) on (1).
\\
        & Case: $\rn{s-start}$. Let $t,l'$ be such that:
\\
        & \z $e_1=\process{t}{l'}$; and
\\
        & \z $\sigma_1=\proj{\sigma}{l}$; and
\\
        & \z $ps_1=\proj{ps}{l}+\process{t}{l'}$; and
\\
        & \z Pick $\Sigma_2 = (\sigma,ps+\process{t}{l'})$ and $e_2=e_1$.
\\
        & \z QED
        & by $\rn{s-start}$.
\\
        & Case: $\rn{s-skip}$.
\\
        & \z Pick $\Sigma_2=\Sigma$ and $e_2=\epsilon$.
\\
        & \z QED
        & by $\rn{s-skip}$.
\\
        & \multicolumn{2}{@{}l}{Case: $\rn{s-send}$. Let $t,l',ps_3,ch,v,t'$ be such that:}
\\
  (2)   & \z $\proj{ps}{l}=ps_3\uplus\{\process{t}{l'}\}$; and
\\
        & \z $ps_1=ps_3\uplus\{\process{t'}{l'}\}$; and
\\
        & \z $e_1=\evsend{ch}{v}$; and
\\
        & \z $\sigma_1=\proj{\sigma}{l}$; and
\\
        & \z $run(t)=\opsend{ch}{v}{t'}$; and
\\
        & \z $l'\sqsubseteq\s{label}(ch)$.
\\
        & \z $l'\sqsubseteq l$
        & by (2).
\\
        & \z $ps_3 = \proj{(ps-\process{t}{l'})}{l}$
        & by (2).
\\
        & \z Pick $\Sigma_2 = (\sigma, ps-\process{t}{l'}+\process{t'}{l'})$.
\\
        & \z Pick $e_2=e_1$.
\\
        & \z QED
        & by $\rn{s-send}$.
\\
        & Remaining cases omitted.
\finishproof

\iffalse

\noindent
\textbf{Theorem (Single step Progress Sensitive Non-Interference).}
\\
\startproof
  (1) & If $\lequiv{l}{\Sigma_1}{\Sigma_2}$
\\
  (2) & and $\sstep{e_1}{\Sigma_1}{\Sigma_1'}$,
\\
      & then $\exists \Sigma_2'.\exists e_2.~\sstep{e_2}{\Sigma_2}{\Sigma_2'}$
        and $\lequiv{l}{\Sigma_1'}{\Sigma_2'}$
        and $\lequiv{l}{e_1}{e_2}$.
\finishproof
\\
\textit{Proof.}
\\
\startproof
      & We have $\Sigma_3',e_3$ such that:
      & (by Projection 1 on (2))
\\
  (3) & $\sstep{e_3}{\proj{\Sigma_1}{l}}{\Sigma_3'}$; and
\\
      & $\lequiv{l}{\Sigma_1'}{\Sigma_3'}$; and
\\
      & $\lequiv{l}{e_1}{e_3}$.
\\
  (4) & $\sstep{e_3}{\proj{\Sigma_2}{l}}{\Sigma_3'}$; and
      & by substituting (1) in (3).
\\
      & We have $\Sigma_2',e_2$ such that:
      & (by Projection 2 on (4))
\\
      & $\sstep{e_2}{\Sigma_2}{\Sigma_2'}$; and
\\
      & $\lequiv{l}{\Sigma_3'}{\Sigma_2'}$; and
\\
      & $\lequiv{l}{e_3}{e_2}$.
\\
      & QED
      & by transitivity of $\lequiv{l}{}{}$.
\finishproof

\fi